\documentclass[aps,pra,longbibliography,superscriptaddress,amssymb,reprint,10pt]{revtex4-1}

\usepackage{graphicx}
\usepackage{dcolumn}
\usepackage{bm}
\usepackage{color}
\usepackage{amsmath,amsfonts,amssymb}
\usepackage{graphicx}
\usepackage{siunitx}
\usepackage{booktabs,longtable}
\usepackage{natmove}
\usepackage{hyperref}

\newcommand{\abs}[1]{\left|{#1}\right|}

\newcommand{\interpar}[1]{\left({#1}\right)}

\newcommand{\vect}[1]{\mathbf{#1}}
\let\oldcite\cite
\renewcommand*\cite[1]{ \oldcite{#1}}
\DeclareMathOperator{\diver}{div}

\AtBeginDocument{%
	\newwrite\bibnotes
	\def\bibnotesext{Notes.bib}
	\immediate\openout\bibnotes=\jobname\bibnotesext
	\immediate\write\bibnotes{@CONTROL{REVTEX41Control}}
	\immediate\write\bibnotes{@CONTROL{%
	apsrev41Control,author="08",editor="1",pages="1",title="0",year="1"}}
	 \if@filesw
	 \immediate\write\@auxout{\string\citation{apsrev41Control}}%
	\fi
}%

\begin{document}

\author{William Legrand}
\email{william.legrand@cnrs-thales.fr}
\affiliation{Unit\'e Mixte de Physique, CNRS, Thales, Univ.\ Paris-Sud, Universit\'e Paris-Saclay, Palaiseau 91767, France}
\author{Nathan Ronceray}
\affiliation{Unit\'e Mixte de Physique, CNRS, Thales, Univ.\ Paris-Sud, Universit\'e Paris-Saclay, Palaiseau 91767, France}
\affiliation{\'Ecole Polytechnique, Palaiseau 91128, France}
\author{Nicolas Reyren}
\author{Davide Maccariello}
\author{Vincent Cros}
\email{vincent.cros@cnrs-thales.fr}
\author{Albert Fert}
\affiliation{Unit\'e Mixte de Physique, CNRS, Thales, Univ.\ Paris-Sud, Universit\'e Paris-Saclay, Palaiseau 91767, France}
\title{Modeling the shape of axisymmetric skyrmions in magnetic multilayers}

\date{\today}

\begin{abstract}
We present a comprehensive micromagnetic model of isolated axisymmetric skyrmions in magnetic multilayers with perpendicular anisotropy. Most notably, the essential role of the internal dipolar field is extensively considered with a minimum amount of assumptions on the magnetization profiles. The tri-dimensional structure of the multilayered skyrmions is modeled by their radial profiles in each layer. We first compare the results of the model against a full micromagnetic description in Cartesian coordinates. Our model combines information on both layer-dependent size and chirality of the skyrmions. We also provide a convenient criterion in order to characterize the stability of skyrmions against anisotropic elongations that would break their cylindrical symmetry, which allows to confirm the stability of the determined solutions. Because this model is able to treat magnetization configurations twisted through the thickness of multilayered skyrmions, it can provide predictions on any potential hybrid chirality in skyrmions due to the interplay of Dzyaloshinskii-Moriya and dipolar interactions in multilayers. We finally apply the results of our model to the description of the current-driven dynamics of hybrid chiral skyrmions. Using the Thiele formalism, we show that we can predict the forces exerted on the multilayered skyrmions by vertical spin-polarized currents, which provides a method to conform hybrid skyrmion chiralities and spin-current injection geometries in order to optimize skyrmion motion in multilayers, to the aim of maximizing the current-induced velocity, or canceling the skyrmion Hall angle.
\end{abstract}

\maketitle

\section{Introduction}

\begin{figure}
\includegraphics[width=3.375in, trim= 0cm 0cm 0cm 0cm]{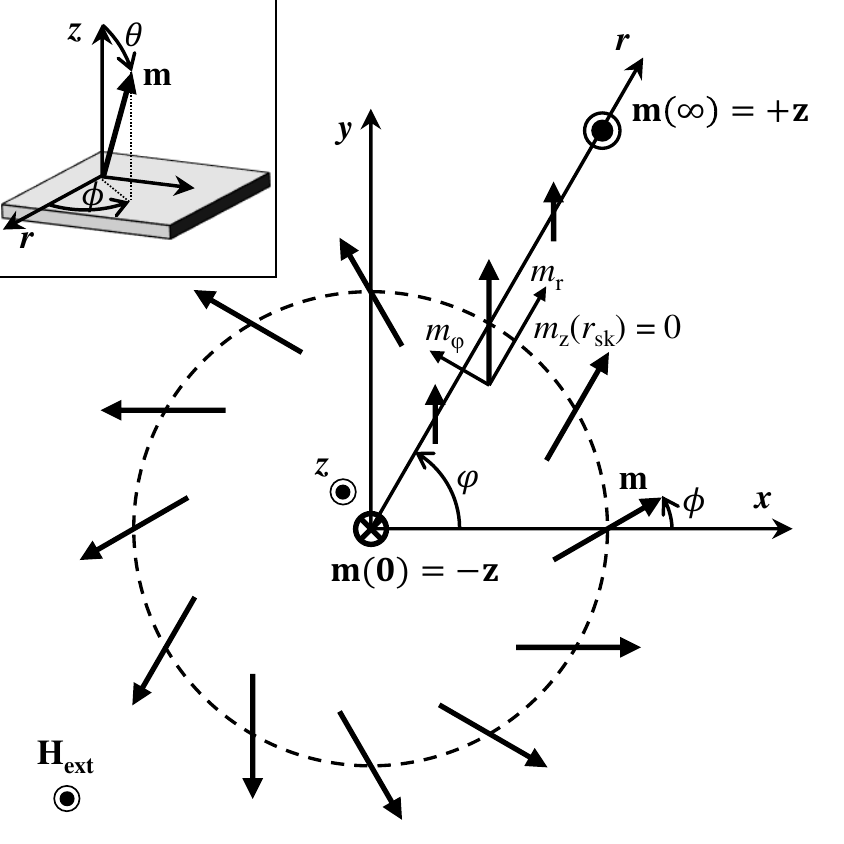}
\caption{Top view of the magnetization in a single ferromagnetic layer hosting a magnetic skyrmion. Thick arrows denotes the reduced magnetization $\vect{m}=\vect{M}/M_{\rm S}$, represented along the radial axis $r$ and around a perimeter of the skyrmion. The axisymmetric skyrmion is fully described by its magnetization components $\interpar{m_{\rm{r}}(r),m_{\rm{\varphi}}(r),m_{\rm{z}}(r)}$, identical along any radial direction, where $\varphi$ is defined as the angle between $x$ and $r$ directions. Here the core magnetization is $\vect{m}(r=0)=-\vect{z}$ and the surrounding magnetization is $\vect{m}(r=+\infty)=+\vect{z}$. The external field $\vect{H_{\rm{ext}}}$ is applied along the $z$ direction. Inset: Definition of magnetization angles, with $\theta$ the polar angle and $\phi$ the azimuthal angle.}
\label{fig:Scheme}
\end{figure}

In the recent years, the discovery of Dzyaloshinskii-Moriya interaction (DMI), not only driven by crystalline order\cite{Dzyaloshinskii1958,Moriya1960} but also by interface inversion asymmetry\cite{Bode2007,Heide2008}, increased the interest for non-collinear magnetic configurations in thin magnetic films and multilayers. Beyond the ferromagnetic and antiferromagnetic configurations, the action of the DMI results in the stabilization of non-uniform configurations where the magnetic order is rotating in one or several directions of the film  plane. Among these configurations, the most studied over these last couple years may probably be the magnetic skyrmions\cite{Bogdanov2001,Rossler2006}, either under the form of isolated skyrmions\cite{Chen2015,Moreau-Luchaire2016,Woo2016} or skyrmion lattices\cite{Muhlbauer2009,Yu2010a}. In a magnetic skyrmion, the magnetization vector in the structure actually maps all directions, resulting in a swirling arrangement, where the magnetization in the center of the structure (the core of the skyrmion) is the opposite of the magnetization in its surrounding environment (see Fig.\ \ref{fig:Scheme}). Moreover, due to the antisymmetric form of the DMI, a well-determined chirality is expected to emerge from these skyrmion configurations, which is determined by the direction and sign of the DMI vector, resulting in a unique rotational sense for the magnetization. This fixed chirality sets a unique, topologically non-trivial configuration for the magnetic order, which results in chirality-related and topology-related effects both in the dynamics of skyrmions (efficient spin-current induced motion\cite{Sampaio2013,Woo2016}, skyrmion Hall effect\cite{Nagaosa2013,Yu2016,Jiang2017,Litzius2017}) and in the transport properties of skyrmionic systems (topological Hall effect\cite{Neubauer2009,Kanazawa2011,Nagaosa2013}). However, in several recent works it has been shown that the particular chirality of the skyrmions set by the DMI could be partially or completely canceled out by competing magnetic interactions, such as the dipolar interactions between magnetic moments, resulting in such cases in three-dimensional arrangements with complex thickness-dependent chiral spin textures\cite{Dovzhenko2016arXiv,Montoya2017,Legrand2018}.

In epitaxially grown ultrathin, single magnetic layers hosting skyrmions, the thickness of the ferromagnetic material is usually one or a couple atomic layers\cite{Romming2013,Romming2015}. In these systems, the thickness of the ferromagnetic material $t$ is then much less than the characteristic dipolar length $l_{\rm{dip}}=\sigma/(\mu_0M_{\rm{S}}^2)$, where $\sigma$ is the domain wall energy density, and $M_{\rm{S}}$ the saturation magnetization. For this reason, it is adequate to neglect the long-range effects of the dipolar field. The dipolar interactions are then equivalent to a reduction of the out-of-plane uniaxial magnetic anisotropy $K_{\rm{u}}$ to an effective magnetic anisotropy $K_{\rm{eff}}=K_{\rm{u}}-\mu_0M_{\rm{S}}^2/2$. In this case, the magnetization texture and the unique chirality set by the DMI remain largely unaffected by the dipolar interactions.

In slightly thicker magnetic layers\cite{Jiang2015, Boulle2016}, needed to stabilize skyrmions up to room-temperature, the requirement of keeping a perpendicular magnetic anisotropy in spite of the mainly interfacial origin of this anisotropy limits the thickness of the ferromagnetic layer to at most \SI{2}{\nano\meter}. Nevertheless, such thicknesses are already large enough for the dipolar interactions to play a critical role in the stabilization of the skyrmions. As the domain wall energy density $\sigma=4\sqrt{AK_{\rm{eff}}}-\pi{}D$ (where $A$ is the stabdard exchange stiffness, and $D$ is the DMI magnitude) is drastically reduced in the presence of the DMI that stabilizes the skyrmions, $t\approx{}l_{\rm{dip}}$ and the dipolar energies are no longer negligible\cite{Boulle2016,Bernand-Mantel2018,Hrabec2017a}. In all practical cases, even if skyrmion energies and sizes are significantly affected, the DMI is still strong enough to ensure a unique chirality of the skyrmions. However, in such single magnetic layers the thermal stability of the skyrmions often remains too weak, so that undesired fluctuation of the skyrmion position\cite{Tolley2018} and/or spontaneous creation or annihilation of skyrmions\cite{Yu2016} have been observed, which imposes to find solutions to improve their room-temperature stability.

A very efficient approach in order to enhance the room-temperature stability of magnetic skyrmions is to stack several ferromagnetic layers by repeating an asymmetric combination of three or four layers, including also heavy-metal layers and spacers in contact with the ferromagnetic layers\cite{Moreau-Luchaire2016,Woo2016,Soumyanarayanan2017,Legrand2017,Litzius2017}. In this way, the effective magnetic volume of the skyrmions is increased, without affecting the DMI and the perpendicular magnetic anisotropy in each single layer, which allows skyrmions to better resist thermal fluctuations. Such skyrmions in multilayered systems may be called columnar skyrmions, as they are actually made of a vertically aligned stacking of coupled skyrmions hosted by each individual ferromagnetic layer. In many-repeats multilayers or for large saturation magnetization $M_{\rm{S}}$ values, a very good thermal stability can be achieved even at room-temperature. However, the counterpart is that the dipolar field becomes very significant in these systems, not only modifying the energies of the skyrmions but also affecting their internal magnetization texture\cite{Lemesh2017,Buttner2018}. The demagnetizing effects of the dipolar field are predicted to cause a rotation of the in-plane magnetic moments of the skyrmions, driving a reorientation from N\'eel skyrmions, as favored by the form of the interfacial DMI, into Bloch skyrmions, as favored by the dipolar interactions. In such multilayers, a right balance between enough layers to reach sufficient thermal stability, but few enough layers to mitigate the reorientation effect of dipolar interactions, needs to be found if a N\'eel configuration is necessary\cite{Buttner2018}.

Beyond such N\'eel to Bloch transition effect, we have demonstrated experimentally in a previous work\cite{Legrand2018} that despite the presence of a large DMI in Pt/Co/AlO$_{\rm{x}}$ based magnetic multilayers, the in-plane magnetic moments of the skyrmions are actually reversed by the dipolar interactions in some, but not all, of the magnetic layers, resulting in the stabilization of hybrid chiral structures. It is thus to be emphasized that above a critical number of repetitions of the magnetic layers, the chirality of the skyrmions thus varies across the thickness of the multilayers\cite{Legrand2018} instead of reorienting coherently through the layers. The magnetization texture is thus no longer $z$-independent, contrary to what has been often hypothesized. Because the current-driven dynamics of skyrmions are related to the details of their internal magnetization texture\cite{Sampaio2013}, such alterations of the chirality are important and require particular attention, in order to understand how to efficiently manipulate such skyrmions with currents\cite{Woo2016,Woo2017,Litzius2017,Fert2017}. However, to date, most attempts in order to model the profile of magnetic skyrmions consider a uniform magnetization across the thickness of the stack and have neglected the layer-by-layer variations of the magnetic configuration.

It is the central objective of this article to tackle these issues, by providing a numerical model that is able to predict the actual three-dimensional equilibrium profiles of axisymmetric skyrmions in magnetic multilayers. Our motivation is to take advantage of the rotational symmetry of the skyrmions for in-plane-isotropic multilayers, in order to get a simple and fast determination of the solution of energy minimization problem by reducing a tri-dimensional problem into a bi-dimensional one. This technique has already been applied to the study of magnetic bubbles in thick ferromagnets\cite{Tu1971,DeBonte1973}. Here, we extend this approach by including the DMI, the presence of chiral spin textures and the fact that we consider magnetic multilayers. Despite the computational complexity of quantitatively determining the dipolar field without further simplifying assumptions, we find that by making use of their cylindrical symmetry the equilibrium profile of multilayered skyrmions can be determined in reasonable time on conventional personal computers. We first validate our model by comparing the obtained profiles with reference solutions provided by the usual and multi-purpose micromagnetic solver Mumax$^3$, which uses Cartesian, tri-dimensional coordinates. Our method provides us with a convenient tool that we apply, as an example, to the problem of the determination of the size of skyrmions as a function of applied field and number of layers. We also introduce a similar model for double domain walls, in Cartesian coordinates, in order to quantify the stability of the skyrmion solutions against anisotropic deformations. We then rely on the main advantage of the present model, being able to determine the layer-by-layer chirality of the skyrmions, to analyze hybrid chirality in multilayered skyrmions. In turn, this allows us to predict the current-induced dynamics of these skyrmions from their equilibrium configuration, by determining the different terms of the Thiele equation. We finally provide strategies to improve the skyrmion velocity and control their direction of motion.

\section{Description of the model}
\label{sec:model}

\subsection{Magnetic energy terms}

In ferromagnetic thin films, five interaction terms are usually considered in the energy of the magnetic configurations: Heisenberg exchange, Dzyaloshinskii-Moriya interaction, magnetic anisotropy, Zeeman interaction, and long-range dipolar interactions. In the following of this article, to simplify the discussions we will only consider a DMI vector favoring N\'eel walls or N\'eel skyrmions (DMI vector $\vect{D}$ in the plane of the films and perpendicular to the vector joining the related magnetic moments), as it is most often found in metallic multilayers. Note however that our model can easily be adapted to other DMI vectors, adjusting the corresponding term in the following Eq.\,\eqref{eq:Esk}. In isotropic materials for which the DMI thus only has an interfacial origin, the stabilized skyrmions are axisymmetric and their energy in a ferromagnetic layer can be written as\cite{Leonov2016}
\begin{equation}
\label{eq:Esk}
\begin{aligned}
E_{\rm{sk}}=2\pi t\int_{r=0}^{\infty}& \left\{A\interpar{\nabla\vect{m}}^2 + D\interpar{m_{\rm{z}}\diver{\vect{m}}-\vect{m}\nabla{m_{\rm{z}}}}\right. \\
& + K_{\rm{u}} \interpar{1-m_{\rm{z}}^2} + \mu_0H_{\rm{ext}}M_{\rm{S}}\interpar{1-m_{\rm{z}}} \\
&\left. - \mu_0M_{\rm{S}}/2\interpar{M_{\rm{S}}+\vect{H_{\rm{dip}}}\cdot\vect{m}}\vphantom{\interpar{\nabla\vect{m}}^2}\right\} r dr
\end{aligned}
\end{equation}
where $t$ is the thickness of a single ferromagnetic layer of magnetization $M_{\rm{S}}$, $r$ is the radial distance from the center of the skyrmion, $A$ is the exchange stiffness parameter, $D$ is the DMI parameter, $K_{\rm{u}}$ is the out-of-plane uniaxial anisotropy parameter, $H_{\rm{ext}}$ is the external field (here applied perpendicular to the plane) and $\vect{H_{\rm{dip}}}$ is the dipolar field generated by the magnetization distribution. For thin magnetic layers, say, $t\ll{}l_{\rm{dip}}$, we can consider that the magnetization does not vary along $z$ inside a given layer. We take into account the three components of the magnetization profile at any $r$: $m_{\rm{r}}(r)$ (along the radial direction), $m_{\rm{\varphi}}(r)$ (perpendicular to the radial direction) and $m_{\rm{z}}(r)$ (perpendicular to the plane of the layer), as represented in Fig.\ \ref{fig:Scheme}. By introducing the dimensionless radius $\rho=r/\sqrt{A/K_{\rm{eff}}}$, we get a dimensionless energy integral\cite{Knoester2014}
\begin{widetext}
\begin{multline}
\label{eq:Eskred}
E_{\rm{sk}}=2\pi t A\int_{\rho=0}^{\infty} \left\{\frac{1-m_{\rm{z}}^2}{\rho^2}+\interpar{\frac{dm_{\rm{r}}}{d\rho}}^2+\interpar{\frac{dm_{\rm{\varphi}}}{d\rho}}^2+ \interpar{\frac{dm_{\rm{z}}}{d\rho}}^2 +\frac{4D}{\pi D_{\rm{c}}}\interpar{\frac{m_{\rm{r}}m_{\rm{z}}}{\rho}+m_{\rm{z}}\frac{dm_{\rm{r}}}{d\rho}-m_{\rm{r}}\frac{dm_{\rm{z}}}{d\rho}}\right. \\
+\left.\frac{K_{\rm{u}}}{K_{\rm{eff}}}\interpar{1-m_{\rm{z}}^2} + \frac{\mu_0H_{\rm{ext}}M_{\rm{S}}}{K_{\rm{eff}}}\interpar{1-m_{\rm{z}}} - \frac{\mu_0M_{\rm{S}}^2}{2K_{\rm{eff}}}\interpar{1+\frac{\vect{H_{\rm{dip}}}}{M_{\rm{S}}}\cdot\vect{m}}\right\} \rho d\rho \quad ,
\end{multline}
\end{widetext}
where we introduce the critical DMI value for the onset of spin-spirals\cite{Rohart2013} $D_{\rm{c}}=4\sqrt{AK_{\rm{eff}}}/\pi$. The total energy $E_{\rm{tot}}$ is the sum of $E_{\rm{sk}}$ in all layers.

As we will describe below, the effective fields $\vect{H}_{\rm{eff}}=(H_{\rm{r}},H_{\rm{\varphi}},H_{\rm{z}})$ associated to each energy term can also be expressed from $\vect{m}(r)$. Therefore, we decide to find the equilibrium configuration $\vect{m}(r)$ that minimizes $E_{\rm{tot}}$ by the quasi-static time-evolution of the magnetic texture under these fields as obtained from the Landau-Lifshitz equation. To this effect, we only consider the damping term and not the precession term of the Landau-Lifshitz equation (case of a very large damping). After initializing the system with a given configuration $\vect{m}(r)$, the system relaxes directly to the closest state of minimum energy, following the direction given by the sum of all effective fields. To perform such minimization, at each iteration, we thus determine the step increment $\delta_{\vect{m}}$ representing the evolution of $\vect{m}$, obtained from the Landau-Lifshitz equation as
\begin{equation}
\begin{split}
\delta_{\vect{m}} & = -\lambda \; \vect{m}\times\interpar{\vect{m}\times\vect{H_{\rm{eff}}}} \\
 &= \lambda 
\begin{bmatrix}
H_{\rm{r}}\interpar{1-m_{\rm{r}}^2}-H_{\rm{\varphi}}m_{\rm{r}}m_{\rm{\varphi}}-H_{\rm{z}}m_{\rm{r}}m_{\rm{z}}\\
H_{\rm{\varphi}}\interpar{1-m_{\rm{\varphi}}^2}-H_{\rm{z}}m_{\rm{\varphi}}m_{\rm{z}}-H_{\rm{r}}m_{\rm{\varphi}}m_{\rm{r}}\\
H_{\rm{z}}\interpar{1-m_{\rm{z}}^2}-H_{\rm{r}}m_{\rm{z}}m_{\rm{r}}-H_{\rm{\varphi}}m_{\rm{z}}m_{\rm{\varphi}}
\end{bmatrix}\ ,
\end{split}
\end{equation}
where $\lambda$ is a constant small enough to ensure convergence. By adding successively the step increments $\delta_{\vect{m}}$ to $\vect{m}$, the magnetic configuration converges to the closest minimum of energy and thus minimizes $E_{\rm{tot}}$.

The effective field $\vect{H_{\rm{eff}}}$ is the sum of the five field terms deriving from each energy term. The exchange field is obtained by 
\begin{equation}
\begin{split}
\vect{H_{\rm{A}}} & = \frac{2A}{\mu_0M_{\rm{S}}} \nabla^2\vect{m}\\
 &= \frac{2A}{\mu_0M_{\rm{S}}}
\begin{bmatrix}
\frac{d^2m_{\rm{r}}}{dr^2}+\frac{1}{r}\frac{dm_{\rm{r}}}{dr}-\frac{m_{\rm{r}}}{r^2}\\
\frac{d^2m_{\rm{\varphi}}}{dr^2}+\frac{1}{r}\frac{dm_{\rm{\varphi}}}{dr}-\frac{m_{\rm{\varphi}}}{r^2}\\
\frac{d^2m_{\rm{z}}}{dr^2}+\frac{1}{r}\frac{dm_{\rm{z}}}{dr}
\end{bmatrix},
\end{split}
\end{equation}
the DMI field is obtained by 
\begin{equation}
\begin{split}
\vect{H_{\rm{D}}} & = \frac{2D}{\mu_0M_{\rm{S}}} \left[\interpar{\diver{\vect{m}}}\vect{z}-\nabla{m_{\rm{z}}}\right]\\
 &= \frac{2D}{\mu_0M_{\rm{S}}} \left[ \frac{dm_{\rm{z}}}{dr} \vect{r}-\interpar{\frac{m_{\rm{r}}}{r}+\frac{dm_{\rm{r}}}{dr}}\vect{z}\right],
\end{split}
\end{equation}
and anisotropy and external fields are given by $\vect{H_{\rm{K}}}=2K_{\rm{u}}m_{\rm{z}}/(\mu_0M_{\rm{S}})\vect{z}$ and $H_{\rm{ext}}\vect{z}$, respectively. These four field terms are straightforward to determine because they are functions of the local magnetization and its derivatives only. However, most of the complexity in the determination of the skyrmion profiles resides in the determination of the dipolar field $\vect{H_{\rm{dip}}}$. As dipolar interactions constitute a long-range interaction, $\vect{H_{\rm{dip}}}$ at each point is a function of the magnetization at every point in the system. In the present model, the dipolar field is also the only term that couples the different magnetic layers. Following the approach that has been developed for the study of magnetic bubbles in thick ferromagnetic layers, we can find the solution of the magnetostatic problem relying on the cylindrical symmetry of skyrmions.

\subsection{Solution for the dipolar field}
\label{subsec:dipolarsol}

The dipolar field $\vect{H_{\rm{dip}}}$ is defined as the opposite of the gradient of the magnetostatic potential $\psi$, which satisfies Poisson's equation and specific boundary conditions, related to volume and surface magnetic charges, respectively\cite{Tu1971,DeBonte1973}:
\begin{equation}
\label{eq:Poisson}
\begin{aligned}
	\text{(i)}\: & \nabla^2\psi = 
	\begin{cases}
	M_{\rm{S}} \diver{\vect{m}},& -t/2\leq z \leq t/2 \\
	0,& \text{outside}
	\end{cases} \\
	\text{(ii)}\: & \nabla{\psi}\xrightarrow[z \to \pm\infty]{} 0\\
	\text{(iii)}\: & \text{$\psi$ continuous at $z=\pm t/2$} \\
	\text{(iv)}\: & \left.\frac{\partial\psi}{\partial z}\right|_{z=\pm\frac{t}{2}^-} \pm \interpar{-M_{\rm{S}}\vect{m}\cdot\vect{z}} = \left.\frac{\partial\psi}{\partial z}\right|_{z=\pm\frac{t}{2}^+}
\end{aligned}
\end{equation}
where the ferromagnetic layer that is the source $\vect{m}(r)$ of the dipolar field extends between $z=-t/2$ and $z=t/2$. In the following, we apply the principle of superposition used in magnetostatic problems and neglect the interactions between volume charges in Eq.\,\eqref{eq:Poisson}(i) ($f(r)=M_{\rm{S}} \diver{\vect{m}}$) and surface charges in Eq.\,\eqref{eq:Poisson}(iv) ($h(r)=-M_{\rm{S}}\vect{m}\cdot\vect{z}$). This allows us to find the potentials associated to $f$ and $h$ separately. We thus define $\psi_{\rm{f}}$ the potential associated to $f$ with $h=0$, and $\psi_{\rm{h}}$ the potential associated to $h$ with $f=0$.

The resolution of Poisson's equation in cylindrical symmetry uses Hankel transforms, the equivalent of Fourier transform for axisymmetric functions. We choose the Bessel function of the first kind, order 0, $J_0$ and thus define for any function $g(r)$ its Hankel transform
\begin{equation}
\bar{g}(k)=\int_{r=0}^{\infty}rJ_0(kr)g(r)dr
\end{equation}
For the magnetic potential, which satisfies $\Delta\psi=f\> \text{or}\> 0$, its Hankel transform $\bar{\psi}(k,z)$ then verifies for any $k,z$
\begin{equation}
\frac{\partial^2\bar{\psi}(k,z)}{\partial z^2}-k^2\bar{\psi}(k,z)=\bar{f}\> \text{or}\> 0
\end{equation}
which allows to find $\bar{\psi}$ by solving the partial differential equation. The detail of the resolution is given in Appendix \ref{app:Solution_poisson}. We get for the potentials originating from the volumes charges $f$ and and the surface charges $h$
\begin{equation}
\label{eq:psibarsol}
\begin{aligned}
	\bar{\psi_{\rm{f}}}&=\left\{
	\begin{aligned}
	&\frac{\bar{f}}{k^2}\exp{\interpar{-kt/2}}\cosh{\interpar{kz}}-\frac{\bar{f}}{k^2},& 0\leq z \leq t/2 \\
	&\frac{-\bar{f}}{k^2}\sinh{\interpar{kt/2}}\exp{\interpar{-kz}},& z>t/2
	\end{aligned}\right. \\
	\bar{\psi_{\rm{h}}}&=\left\{
	\begin{aligned}
	&\frac{-\bar{h}}{k}\exp{\interpar{-kt/2}}\sinh{\interpar{kz}},& 0\leq z \leq t/2 \\
	&\frac{-\bar{h}}{k}\sinh{\interpar{kt/2}}\exp{\interpar{-kz}},& z>t/2
	\end{aligned}\right.
\end{aligned}
\end{equation}
that can be completed by symmetry. As volume charges are symmetric with respect to $z$ while surface charges are antisymmetric, we have that 
\begin{equation}
\begin{aligned}
\bar{\psi_{\rm{f}}}(k,z)&=+\bar{\psi_{\rm{f}}}(k,-z) \\
\bar{\psi_{\rm{h}}}(k,z)&=-\bar{\psi_{\rm{h}}}(k,-z).
\end{aligned}
\end{equation}
The fields acting on a given layer $l_j$ at position $z'$ (either the source layer $l_i$ itself or another layer) along $r$ and $z$ are finally obtained by transforming back $\bar{\psi}_{\rm{f,h}}$ into $\psi_{\rm{f,h}}$ using the inverse Hankel transform, and then averaging $\partial\psi_{\rm{f,h}}/\partial r$ and $\partial\psi_{\rm{f,h}}/\partial z$ over the thickness $t'$ of the affected layer $l_j$
\begin{equation}
\label{eq:Hdip}
\begin{aligned}
H_{\rm{dip,r}}^{i,j}(r)=&-\frac{1}{t'}\int_{z=z'-t'/2}^{z'+t'/2} \interpar{\frac{\partial\psi_{\rm{f}}}{\partial r}+\frac{\partial\psi_{\rm{h}}}{\partial r}}dz \\
H_{\rm{dip,z}}^{i,j}(r)=&\frac{\left.\interpar{\psi_{\rm{f}}+\psi_{\rm{h}}}\right|_{z'-t'/2}-\left.\interpar{\psi_{\rm{f}}+\psi_{\rm{h}}}\right|_{z'+t'/2}}{t'}.
\end{aligned}
\end{equation}
Thus, $\vect{H}_{\rm{dip}}(\vect{m_i}(r),t,t',\Delta z)$ is defined as a function of source magnetization configuration $\vect{m_i}(r)$ and layer thickness $t=t_i$, affected layer thickness $t'=t_j$ and interlayer spacing $\Delta z$. Finally, for a multilayer comprising $L$ layers, for any layer $l_j$ we obtain the total dipolar field
\begin{equation}
\vect{H}_{\rm{dip}}^{\rm{tot,j}}= \vect{H_{\rm{dip}}}(\vect{m_j},t_j,t_j,0) + \sum_{i=1,i\neq{}j}^L \vect{H_{\rm{dip}}}(\vect{m_i},t_i,t_j,z_j-z_i)
\end{equation}
where we separate self-interacting term and interlayer interactions term, as they have different forms from Eq.\,\eqref{eq:psibarsol}.

With the model of magnetic interactions in multilayered skymion systems that we describe here, a tri-dimensional problem is thus reduced into a bi-dimensional problem, as the geometric variables are layers $l_1,l_2,\dots,l_L$ and radius, which can be discretized in points $r_1,r_2,\dots,r_N$. It allows the study of the profiles of multilayered skyrmions by direct numerical minimization. As we demonstrate below, it offers an efficient tool for understanding different issues related to the chirality of skyrmions in such multilayered systems. In order to focus on the presentation of the model without going too far into its technical aspects, we have actually skipped several details regarding its numerical implementation. However, we believe that it can be of interest to discuss some points of the implementation, as they are essential to get an accurate determination of the energy and a fast convergence of the model. An extended description of the implementation of the model is thus given in Appendix \ref{app:Condnum}.

\section{Validation of the model: comparison with Mumax$^3$}
\label{sec:validation}

\begin{figure}
\includegraphics[width=3.375in, trim= 0cm 0cm 0cm 0cm]{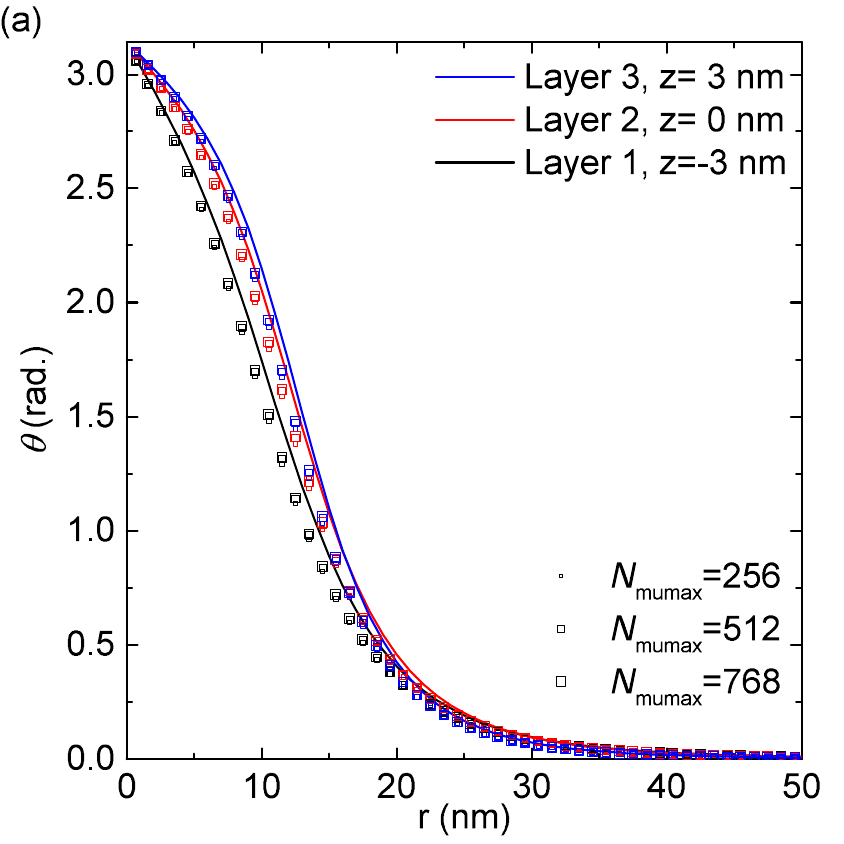}
\includegraphics[width=3.375in, trim= 0cm 0cm 0cm 0cm]{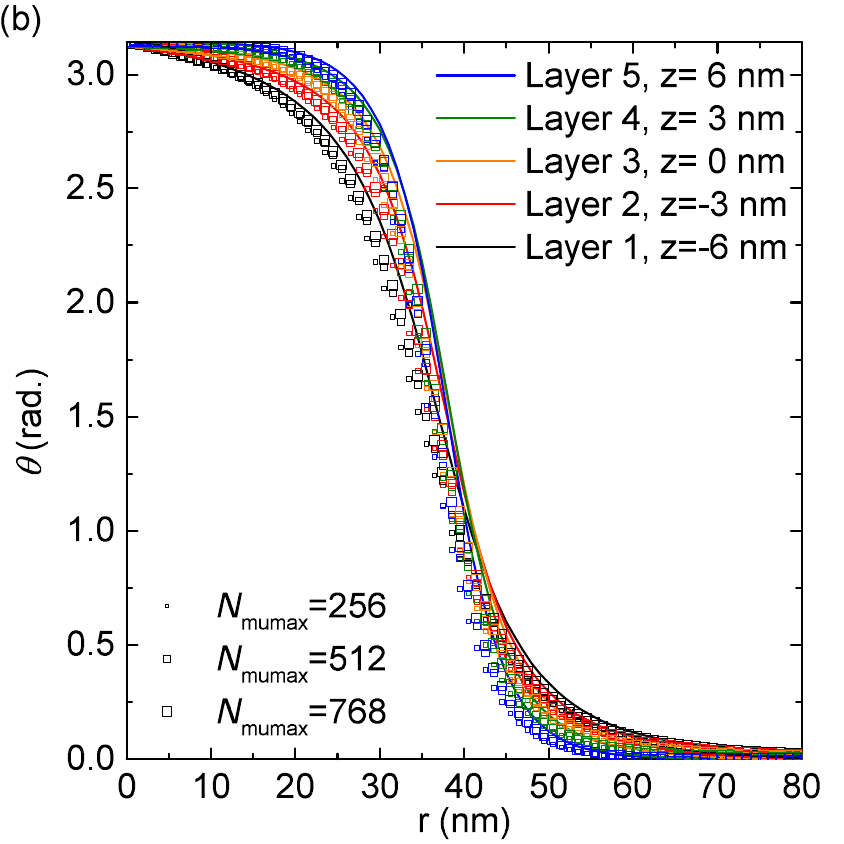}
\caption{Multilayer model comparison for (a) $L=3$ and (b) $L=5$ layers. Colored lines are the $\theta(r)$ profiles in each layer from our $\vect{m}(r)$ model, while hollow squares are the results from Mumax$^3$ package. The increasing sizes of the hollow squares correspond to the cases $N_{\rm{mumax}}=256, 512, 768$.}
\label{fig:Comparemumax}
\end{figure}

Before using the present model to analyze the properties of isolated skyrmions in multilayers, we check the validity of our $\vect{m}(r)$ solver by comparing its results with the standard micromagnetic simulation package\cite{Vansteenkiste2014} Mumax$^3$. In the following, we will always consider multilayers with fixed layer thickness ($t_i=t_{\rm{FM}}$ for all layers) and periodicity $p$ between layers, even if we note that the model we have presented above can also be used to model the case of arbitrary layer thicknesses and positions. We select a set of standard parameters for Co based multilayers\cite{Moreau-Luchaire2016} : $A =$ \SI{10}{\pico\joule\per\meter}, $D =$ \SI{1.35}{\milli\joule\per\meter\squared}, $M_{\rm{S}} =$ \SI{1}{\mega\ampere\per\meter}, $K_{\rm{u}} =$ \SI{0.8}{\mega\joule\per\meter\cubed} (which corresponds to $K_{\rm{eff}} =$ \SI{0.172}{\mega\joule\per\meter\cubed}), $\mu_0H_{\rm{ext}} =$ \SI{50}{\milli\tesla},  $t_{\rm{FM}} = $ \SI{1}{\nano\meter} and $p = $ \SI{3}{\nano\meter}. We initialize the magnetization in all layers with a usual approximation\cite{Romming2015} of the skyrmion profile
\begin{equation}
\begin{aligned}
\vect{m}=\interpar{\sin(\theta)\cos(\phi),\sin(\theta)\sin(\phi),\cos(\theta)} \\
\theta(r)=\frac{\pi}{2} - \arcsin{\left[\tanh{\interpar{\frac{r-4\delta}{\delta}}}\right]}, \phi(r)=\pi/4,
\end{aligned}
\end{equation}
where $\delta=\sqrt{A/K_{\rm{eff}}}$ is the domain-wall width, and where the initial $\phi$ is chosen on-purpose in-between Bloch ($\phi=\pm\pi/2$) and N\'eel ($\phi=0, \pi$) configurations, in order to let the system relax to its most stable state. Note that this initial configuration will determine the Bloch component of the magnetization texture (towards $\phi = \pi/2$) if any, but the solution with an opposite Bloch component (towards $\phi = -\pi/2$) is equally valid, because no interaction or sample geometry breaks this symmetry in the absence of bulk DMI. We let both algorithms (our solver and Mumax$^3$) converge to $\abs{\delta_{\vect{m}}}<10^{-9}$ and compare the resulting profiles.

Before we analyze the results, we have to note that both solvers actually treat two different problems. In our $\vect{m}(r)$ solver, we study an isolated skyrmion within an infinite, uniformly magnetized layer, whereas in the case of Mumax$^3$, the simulated system always has a finite size and the magnetization outside it is only mimicked with periodic boundary conditions. To check that both methods agree in the limit of very large simulation sizes in Mumax (number of cells $N_{\rm{mumax}}\to\infty$), we compare our model to different runs of Mumax$^3$ with increasing simulation sizes $N_{\rm{mumax}}=256, 512, 768$. The cell size is fixed at $1\times 1\times 1$ \si{\cubic\nano\meter}. We consider two geometries with $L=3$ and $L=5$ layers, for which we present the resulting skyrmion profiles in Figs.\ \ref{fig:Comparemumax}(a) and \ref{fig:Comparemumax}(b). In both cases, we find a very good agreement between our solver and Mumax$^3$. Due to the large $D$, skyrmions are N\'eel with $\phi=0$ in all layers. As we can see from the plots of layer-by-layer profiles $\theta(r)$ with increasingly large hollow squares, corresponding to increasing sizes of the simulation grid with $N_{\rm{mumax}}=256, 512, 768$, the skyrmion profile in the finite geometry of Mumax progressively converges to our skyrmion profile in the infinite geometry, represented by the lines. We have also checked that the values of the effective fields found by both models are equal for $L=3$ up to $L=20$, and at different values of $\phi(r)$ angles.

\section{Application to the problem of the size of skyrmions}

\begin{figure*}
\includegraphics[width=3.375in, trim= 0cm 0cm 0cm 0cm]{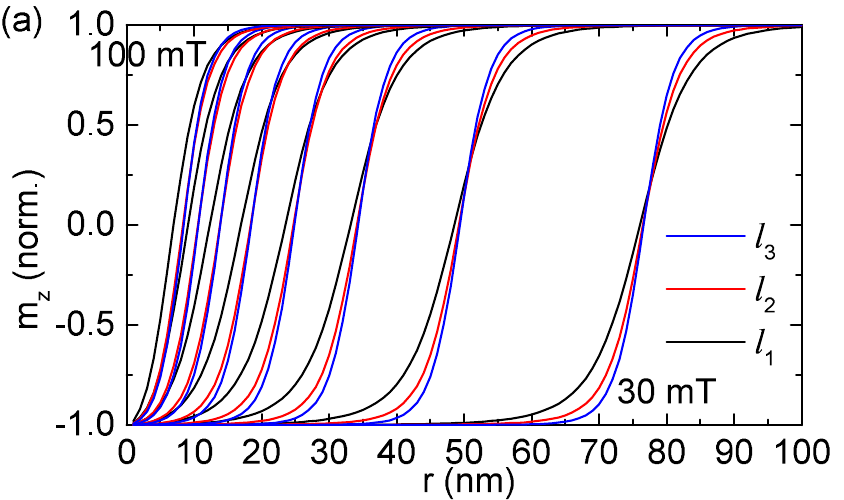}
\includegraphics[width=3.375in, trim= 0cm 0cm 0cm 0cm]{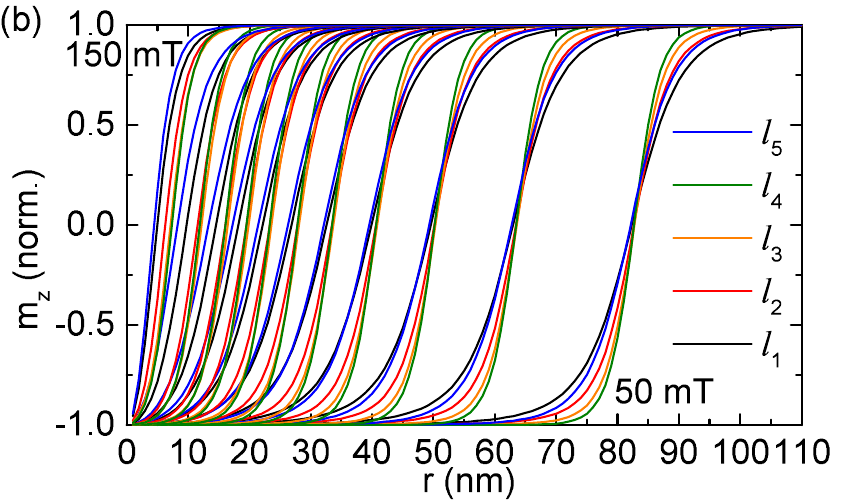}
\includegraphics[width=3.375in, trim= 0cm 0cm 0cm 0cm]{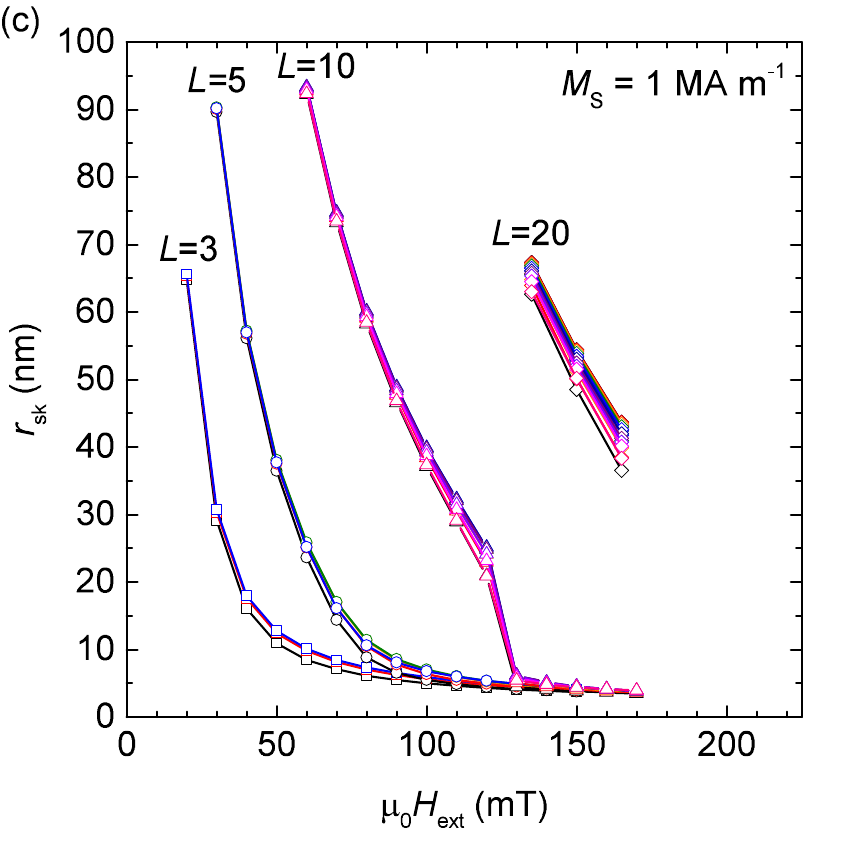}
\includegraphics[width=3.375in, trim= 0cm 0cm 0cm 0cm]{{{Sk_sizes_Ms=1.3e6}}}
\caption{Profile of the skyrmion at different external fields, by \SI{10}{\milli\tesla} increments, for (a) $L=3$ and (b) $L=5$ layers with $M_{\rm{S}} =$ \SI{1.3}{\mega\ampere\per\meter}. The different colors of lines are the $m_{\rm{z}}(r)$ profiles in each layer. Layer-by-layer skyrmion radius $r_{\rm{sk}}$ as a function of external field, for $L=3,5,10,20$, in the cases of (c) $M_{\rm{S}} =$ \SI{1.0}{\mega\ampere\per\meter} and (d) $M_{\rm{S}} =$ \SI{1.3}{\mega\ampere\per\meter}. The different colors distinguish between the different layers.}
\label{fig:Bdepprofiles}
\end{figure*}

In this section, we begin by a study of the equilibrium sizes (and profiles) of the skyrmion in each layer as a function of the external magnetic field $H_{\rm{ext}}$. Issues related to the chirality of skyrmions will be addressed later, in the following sections. We first describe the cases of $L=3$ and $L=5$, for fields varying in the ranges $\mu_0H_{\rm{ext}} =$ 30--\SI{100}{\milli\tesla} and 50--\SI{150}{\milli\tesla}, respectively. Note that in our model the skyrmion is embedded in an infinite plane; in particular there is no confinement due to the problem geometry or edge effects\cite{Rohart2013}. We only consider cases for which the finite simulation space does not influence the profile of skyrmions, that is, $m_{\rm{z}}$ very close to 1 at the maximum $r_N$ considered.

For this first example of application of our model, we use $M_{\rm{S}} =$ \SI{1.3}{\mega\ampere\per\meter}, $K_{\rm{u}} =$ \SI{1.2}{\mega\joule\per\meter\cubed}, and other parameters same as above, which is typical of pure Co- and Fe-based ferromagnetic layers\cite{Soumyanarayanan2017}. The obtained profiles are reported in Figs.\ \ref{fig:Bdepprofiles}(a) and \ref{fig:Bdepprofiles}(b). For increasing values of the external field, we notice an evolution from bubble skyrmions, which exhibit an extended core, at lower fields, towards compact skyrmions, at larger magnetic fields. As we can see from the different shapes of the profiles in each layer, they are differently affected by the external field, even for multilayers restricted to a few layers. To extend this study, we present the evolution of the skyrmion size in each layer as a function of the external field, for both $M_{\rm{S}} =$ \SI{1.0}{\mega\ampere\per\meter}; $K_{\rm{u}} =$ \SI{0.8}{\mega\joule\per\meter\cubed} [Fig.\ \ref{fig:Bdepprofiles}(c)] and $M_{\rm{S}} =$ \SI{1.3}{\mega\ampere\per\meter}; $K_{\rm{u}} =$ \SI{1.2}{\mega\joule\per\meter\cubed} [Fig.\ \ref{fig:Bdepprofiles}(d)], and multilayers with $L=3,5,10,20$. The skyrmion size $r_{\rm{sk}}$ is defined as the (layer-dependent) radius at which $m_{\rm{z}}=0$ (see Fig.\ \ref{fig:Scheme}).

\begin{figure}
\includegraphics[width=3.375in, trim= 0cm 0cm 0cm 0cm]{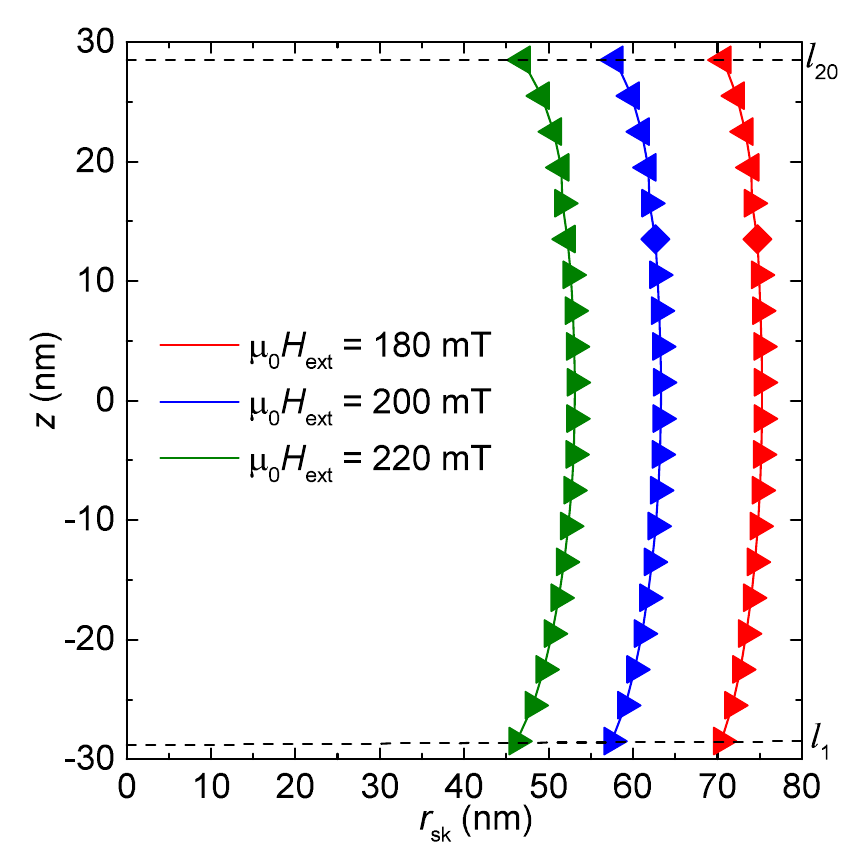}
\caption{Global structure of the skyrmion represented by its layer-dependent $r_{\rm{sk}}$ at different $z$ positions for $L=20$ layers, with $M_{\rm{S}} =$ \SI{1.3}{\mega\ampere\per\meter}, at different external fields $\mu_0H_{\rm{ext}} =$ \SI{180}{\milli\tesla} (red curve), \SI{200}{\milli\tesla}(blue curve), and \SI{220}{\milli\tesla} (green curve). The vertical axis is the $z$ position of the layers from $l_1$ at the bottom to $l_{20}$ on the top. The effective chirality in each layer is represented by the shape of the symbols: right-pointing triangles for N\'eel chirality with $\phi=0$, diamonds for intermediate configurations with $0<\phi<\pi$ and left-pointing triangles for N\'eel chirality with $\phi=\pi$.}
\label{fig:L20profiles}
\end{figure}

As expected, the effects of the dipolar field are more and more pronounced when the number of repeated layers is increased. This appears very clearly in at least three ways. First, at a given radius, the curves of $r_{\rm{sk}}(H_{\rm{ext}})$ are shifted to higher fields for increasing $L$, because more confinement is required to prevent the skyrmion from expanding under the influence of dipolar interactions from more layers. Second, the slope of these curves at a given radius reduces with increasing $L$, which indicates that the external field variations are less significant relative to the intrinsic dipolar field. Third, the split of the $r_{\rm{sk}}(H_{\rm{ext}})$ curves for the different layers increases with $L$ due to the increasing interlayer interactions. These trends are confirmed by comparing the results presented in Figs.\ \ref{fig:Bdepprofiles}(c) and \ref{fig:Bdepprofiles}(d), which shows that increasing $M_{\rm{S}}$ has a similar effect as increasing $L$, as it also results in a stronger dipolar field. For the case of $L=20$, the skyrmion radius can thus vary by more than $20 \%$ between the external and the central layers of the stacking, resulting in a barrel shape for the skyrmion, as evidenced by the profiles in Fig.\ \ref{fig:L20profiles}, corresponding to the case $M_{\rm{S}} =$ \SI{1.3}{\mega\ampere\per\meter}. As a consequence, it appears important to consider the impact of dipolar interactions and of the multilayered nature of such skyrmions in order to properly describe their profiles and sizes.

\section{Stability of isolated skyrmions against anisotropic deformations}

Even if a skyrmion configuration may minimize the energy in the space of axisymmetric solutions, it may not be a minimum of energy if the symmetry constrain is relaxed. In particular, we have to check that the skyrmion solutions that we find are stable against anisotropic deformations\cite{Bogdanov1994a}, or equivalently, that the skyrmion deformation towards a stripe domain is energetically disfavored. To perform this verification, a first possibility is to add small perturbations to $\vect{m}(x,y,z)$ in the full, three-dimensional Cartesian micromagnetic model and let the system evolve, but this is computationally expensive. Here we propose, as an alternative approach, to establish an approximate criterion for this stability by considering the energy of a stripe domain. Let us consider a domain profile $\vect{m}(x,z)$ with $\vect{m}(x,z)=\vect{m}(r,z)$ for $x=r>0$ and its symmetric for $x<0$, {\it i.e.}, the same profile than a cut through the skyrmion core along $x$ direction, but uniform along the $y$ direction. As this profile is suited to the cylindrical geometry of a skyrmion, we have to relax it to its most stable $360^{\circ}$-domain walls form, proceeding with a similar energy minimization procedure as described above, but assuming instead a linear geometry (uniform along $y$). We provide in Appendix \ref{app:Eskcart} the form of the stripe energy per unit length $E_{\rm{st}}$, that we have used to find the stability of previously determined solutions. The total energy $E'_{\rm{tot}}$ is the sum of $E_{\rm{st}}$ in all layers. If the resulting stripe has a total energy lower than the saturated state, the system will decrease its energy through deformation of the skyrmion, splitting it into two half-skyrmions (also called merons) located at the two ends of a stripe domain and thus the skyrmion will be unstable. 

\begin{figure*}
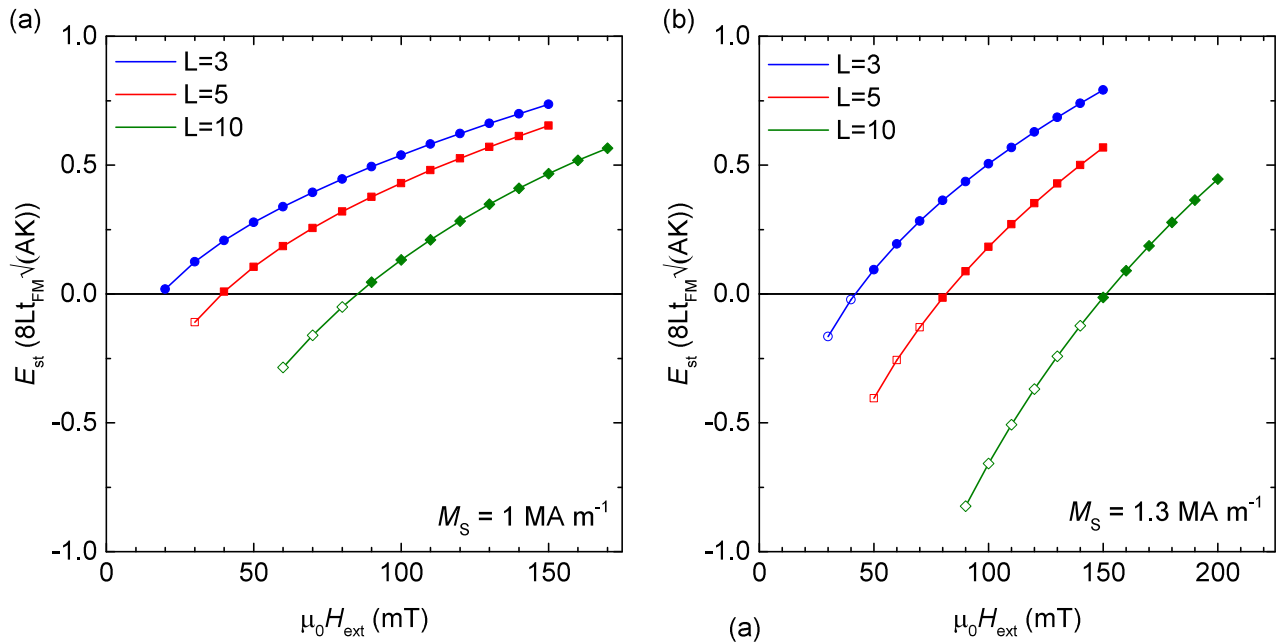

\includegraphics[width=3.375in, trim= 0cm 0cm 0cm 0cm]{{{Sk_stripes_Ms=1e6}}}
\includegraphics[width=3.375in, trim= 0cm 0cm 0cm 0cm]{{{Sk_stripes_Ms=1.3e6}}}
\caption{Energy of the $360^{\circ}$-domain wall stripe as a function of the external field, for $L=3,5,10$, in the cases of (a) $M_{\rm{S}} =$ \SI{1.0}{\mega\ampere\per\meter} and (b) $M_{\rm{S}} =$ \SI{1.3}{\mega\ampere\per\meter}. Circles correspond to $L=3$, squares to $L=5$, and diamonds to $L=10$. The filled symbols correspond to stable skyrmions in full, tri-dimensionnal micromagnetic simulations, while hollow symbols correspond to unstable skyrmions (subject to strip-out instability) in full, tri-dimensionnal micromagnetic simulations.}
\label{fig:Estripe}
\end{figure*}

In Fig.\ \ref{fig:Estripe}, we show the $360^{\circ}$-domain walls energies, obtained using two sets of magnetic parameters as above: $M_{\rm{S}} =$ \SI{1.0}{\mega\ampere\per\meter}; $K_{\rm{u}} =$ \SI{0.8}{\mega\joule\per\meter\cubed} in Fig.\ \ref{fig:Estripe}(a), and $M_{\rm{S}} =$ \SI{1.3}{\mega\ampere\per\meter}; $K_{\rm{u}} =$ \SI{1.2}{\mega\joule\per\meter\cubed}, in Fig.\ \ref{fig:Estripe}(b). We consider three different geometries $L=3$, $L=5$, and $L=10$. As the usual single layer, single domain wall energy per surface unit is $4\sqrt{AK_{\rm{eff}}}$, the stripe energy, without DMI and domain-domain dipolar interactions energy gain, for $L$ layers, is around $8Lt_{\rm{FM}}\sqrt{AK_{\rm{eff}}}$. The stripe energy of the $360^{\circ}$-domain walls $E'_{\rm{tot}}$ is thus evaluated in units of $8Lt_{\rm{FM}}\sqrt{AK_{\rm{eff}}}$. When the DMI and dipolar interactions compensate for this energy cost added to Zeeman energy cost, $E'_{\rm{tot}}<0$ so the stripe configuration is favored over the uniform state and the skyrmion will expand into a stripe domain. We can see that $E'_{\rm{tot}}<0$ only for $\mu_0H_{\rm{ext}} =$ 30--\SI{40}{\milli\tesla} in the case $M_{\rm{S}} =$ \SI{1.3}{\mega\ampere\per\meter}; $K_{\rm{u}} =$ \SI{1.2}{\mega\joule\per\meter\cubed} for $L=3$ and in more cases for $L=5$ and $L=10$. For comparison, the stability as obtained from small perturbations to $\vect{m}(x,y,z)$ in the full, three-dimensional Cartesian micromagnetic model is also reported on these graphs by the interior of the symbols: hollow symbols correspond to unstable skyrmion configurations whereas filled symbols correspond to stable skyrmion configurations. A very good agreement is found, as the unstable configurations from the perturbation method match with negative minimized $E'_{\rm{tot}}$ in all cases tested here except for $M_{\rm{S}} =$ \SI{1.3}{\mega\ampere\per\meter}, with $L=5$ and $\mu_0H_{\rm{ext}} =$ \SI{80}{\milli\tesla}; $L=10$ and $\mu_0H_{\rm{ext}} =$ \SI{150}{\milli\tesla}. In these cases, even if the stripe configuration is more stable, because the energy gain associated to stripes is extremely small, a small energy barrier remains between the skyrmion configuration and the stripe configuration. The skyrmion thus has a very small stability, which would however not resist against thermal fluctuations. Finally, $E'_{\rm{tot}}>0$ is a satisfactory criterion to estimate the skyrmion stability against anisotropic deformations, that is, the validity of the solutions of our axisymmetric model.

However, our model does not provide a direct estimation of the thermal stability\cite{Buttner2018} of the calculated skyrmions states, but only relative to the stripe state. We note that an interesting approach could be to use the calculated profiles to perform nudged elastic band calculations and to determine properly their stability against thermal fluctuations\cite{Rohart2016,vonMalottki2017,Bessarab2018}. 

\section{Chirality: DMI vs.\ Dipolar interactions}

In the previous sections, we have focused on the layer-dependent size of the magnetic skyrmions, which has clearly shown how dipolar interactions affect the skyrmion profiles, and their stability. We will now focus on the actual chirality of the multilayered skyrmions, and will describe how the competition between interfacial DMI and dipolar interactions can result in hybrid chiral textures with different chiralities in the different layers. Importantly, we emphasize that such hybrid chiral configurations can be stabilized even for a limited number of repetitions $L$, as demonstrated hereafter.

\begin{figure*}
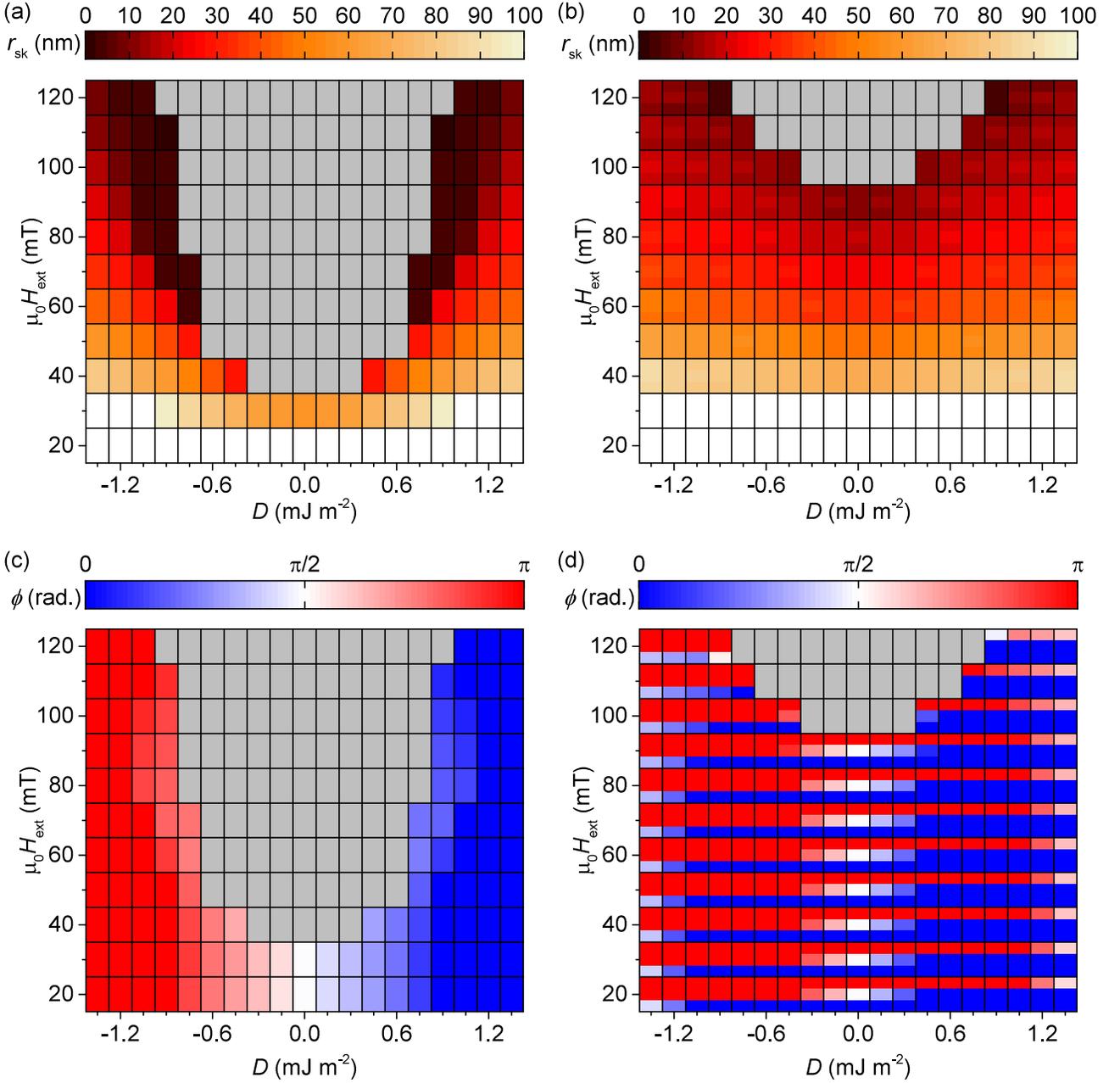

\includegraphics[width=3.375in, trim= 0cm 0cm 0cm 0cm]{{{Sk_sizes_BD_t=1.4e-3_1D}}}
\includegraphics[width=3.375in, trim= 0cm 0cm 0cm 0cm]{{{Sk_sizes_BD_t=1.4e-3_Full}}}
\includegraphics[width=3.375in, trim= 0cm 0cm 0cm 0cm]{{{Sk_chirality_BD_t=1.4e-3_1D}}}
\includegraphics[width=3.375in, trim= 0cm 0cm 0cm 0cm]{{{Sk_chirality_BD_t=1.4e-3_Full}}}
\caption{Equilibrium size of the skyrmion as a function of DMI parameter and external field, for (a) a model with uniform magnetization through the thickness and (b) a layer-by-layer description. The size corresponding to the color in each cell is given by color scale. Skyrmion effective chirality, given by the average in-plane angle of the magnetization with respect to the radial direction, for (c) a model with uniform magnetization through the thickness and (d) a layer-by-layer description. As introduced before, $\phi=0$ means a N\'eel skyrmion with counter-clockwise chirality, $\phi=\pi/2$ means a Bloch skyrmion and $\phi=\pi$ means a N\'eel skyrmion with clockwise chirality, as given by the color scale. The gray color signifies that the skyrmion collapses below the cell size for these parameters. For the layer-by-layer model, inside a cell of the gird which corresponds to a single set of parameters, the three stacked colors from bottom to top correspond to the three layers from bottom to top.}
\label{fig:BDProfiles}
\end{figure*}

We describe the case of a multilayer with $L=3$ and low effective anisotropy $K_{\rm{eff}}$, typical among room-temperature skyrmion systems. In such systems the combination of low $K_{\rm{eff}}$ and non-zero DMI favors the rotation of the magnetization through the plane, which helps to stabilize skyrmions under the bias of the external magnetic field $H_{\rm{ext}}$. The accumulation of three magnetic layers of relatively large thickness increases their stablity. Here we consider $t_{\rm{FM}}= $ \SI{1.4}{\nano\meter} and $p= $ \SI{3.4}{\nano\meter}; $M_{\rm{S}} =$ \SI{1.3}{\mega\ampere\per\meter}; $K_{\rm{u}} =$ \SI{1.2}{\mega\joule\per\meter\cubed}. In order to study quantitatively the competition between dipolar interactions and DMI, we vary the DMI magnitude between $D =$ \SI{-1.35}{\milli\joule\per\meter\squared} and $D =$ \SI{1.35}{\milli\joule\per\meter\squared}, which is the largest effective value of $D$ that has been achieved experimentally\cite{Soumyanarayanan2017} for magnetic layers as thick as \SI{1.4}{\nano\meter}. With this combination of parameters, $\abs{D}<D_{\rm{c}}$ for all values of $D$ considered. In order to highlight the importance of considering a layer-by-layer description for this kind of multilayers, we compare, in Fig.\ \ref{fig:BDProfiles}, the results of our model (called layer-by-layer) to the results of a simpler model (called uniform), in which we impose that all layers share the same profile\cite{Buttner2018}. In each panel, one cell of the grid corresponds to one set of ($D$, $\mu_0H_{\rm{ext}}$) parameters. In the panels of the layer-by-layer description, each cell of the grid actually shows the results of the model for the three different layers.

First, the results for the size of the skyrmion are shown in Figs.\ \ref{fig:BDProfiles}(a) and \ref{fig:BDProfiles}(b) for the uniform model and the layer-by-layer model, respectively. As before, for the layer-by-layer case $r_{\rm{sk}}$ is defined as the (layer-dependent) radius at which $m_{\rm{z}}=0$. In the present case, a very small layer-by-layer variation of the skyrmion size is found in our model, as shown by the almost uniform color in each cell of the panel in Fig.\ \ref{fig:BDProfiles}(b). However, the skyrmion size dependence is much less abrupt in our model than in the uniform model, and leads to less cases for which the skyrmion collapses ($r_{\rm{sk}}$ below the cell size of \SI{1}{\nano\meter}, in gray). This is due to the complete modeling of the interlayer interactions and difference of magnetization profiles between layers. The extra degree of liberty given by the possibility of layer-by-layer variations of the magnetization allows the system energy to be reduced and keeps stable skyrmions in a larger range of parameters. Second, the average value $\left<\phi\right>$ of the in-plane azimuthal angle (later on called effective chirality) of the skyrmion in each layer are shown in Figs.\ \ref{fig:BDProfiles}(c) and \ref{fig:BDProfiles}(d) for the uniform model and the layer-by-layer model, respectively. For the case of the layer-by-layer description, the three layers from bottom to top are represented by the three stacked colors in each cell. Whereas the $D$-dependence of the chirality is, as expected, nearly linear\cite{Thiaville2012,Buttner2018} for the uniform model description [in Fig.\ \ref{fig:BDProfiles}(c), the color indicating the chirality evolves linearly from the center to the sides of the figure], it is no longer linear for the refined layer-by-layer description [in Fig.\ \ref{fig:BDProfiles}(d)]. 

When all layers are considered individually, the effective chiralities of the skyrmion in the different layers thus evolve depending on $D$ through different steps. In the absence of DMI, only dipolar field is present, which is able to reorientate the skyrmions of the bottom and top layers into N\'eel skyrmions of opposite chiralities instead of Bloch skyrmions, while the skyrmion of the central layer remains Bloch [central column of Fig.\ \ref{fig:BDProfiles}(d)]. When a small DMI is progressively added, the chirality of the skyrmion in the central layer progressively evolves towards N\'eel of the chirality determined by the sign of the DMI. For the central layer, this is the same scenario as in the single layer case, where the intralayer dipolar interactions are progressively overcome by the DMI. Finally, to reorientate the last remaining N\'eel skyrmion of chirality opposed to the one favored by DMI (either located in the top or bottom layer depending on the sign of $D$), the interlayer dipolar interactions must also be overcome, rather than only the intralayer dipolar interactions. This shifts the reversal of chirality in the last layer to much larger values of $D$. As can be seen in Fig.\ \ref{fig:BDProfiles}(d), for the lowest field value, $\abs{D}>$ \SI{1.2}{\milli\joule\per\meter\squared} is required for the chirality of the last layer skyrmion to evolve. For $\abs{D}>$ \SI{1.65}{\milli\joule\per\meter\squared} (not shown), a N\'eel skyrmion with a uniform chirality in accordance with the sign of the DMI is recovered. The competition between DMI and dipolar interactions occurs very similarly when $L$ is increased, shifting the reversal of chirality to higher values of $\abs{D}$ in all but the central layer (if any, only for odd values of $L$). As such reorientation effects are due to the dipolar field, an even higher value of $D$ is required to get a uniform chirality for larger values of $M_{\rm{s}}$, for larger layer thicknesses, and for larger values of $L$. Note that the present case has been chosen to highlight the role of the dipolar field, choosing large $M_{\rm S}$ and thick ferromagnetic layers.

For the case of $L=20$ that has been studied in previous sections, the effective chirality in each layer is shown by the shape of the symbols in Fig.\ \ref{fig:L20profiles}. Layers $l_{1}$--$l_{14}$ are N\'eel with $\phi=0$ (right-pointing triangles), layers $l_{17}$--$l_{20}$ are N\'eel with $\phi=\pi$ (left-pointing triangles), and a transition occurs in layers $l_{15}$ and $l_{16}$ (with diamonds for Bloch or partially Bloch). The complex transition, not being always simply N\'eel to Bloch to opposite N\'eel, but being in some cases a succession of alternating chiralities, seems to be a complex consequence of the shape of the dipolar field. It is confirmed by full micromagnetic simulations using Mumax$^3$. Even for a value of $D$ as high as \SI{1.35}{\milli\joule\per\meter\squared}, as in the present case, one fourth of the layers are still hosting N\'eel skyrmions with an effective chirality opposed to the one favored by the DMI, similar to what was found experimentally in a previous work\cite{Legrand2018}.

\section{Application to the current-driven dynamics of hybrid chiral skyrmions}

From the energy minimized $\theta_i(r)$ and $\phi_i(r)$ profiles in all layers obtained above, the expected skyrmion velocities can be directly predicted by making use of the Thiele formalism. In the Thiele approach, the skyrmion is considered as a rigid object, and the global forces acting on the skyrmion can be expressed from the Landau-Lifshitz-Gilbert equation, by integrating the torques acting on the magnetization over the whole magnetization texture\cite{Sampaio2013,Iwasaki2013,Iwasaki2013a}. 
For an axisymmetric skyrmion of profile $\vect{m}=\interpar{m_{\rm{r}},m_{\rm{\varphi}},m_{\rm{z}}}$ we have 
\begin{equation}
\label{eq:ThieleEq}
	\vect{G}\times\vect{v}-\alpha\left[\mathcal{D}\right]\vect{v}+\vect{F}=\vect{0}
\end{equation}
with
\begin{equation}
\begin{aligned}
\label{eq:Thielefactors}
	\vect{G}&=-\frac{M_{\rm{S}}t_{\rm{FM}}}{\gamma}\iint\interpar{\frac{\partial \vect{m}}{\partial x}\times\frac{\partial \vect{m}}{\partial y}}\cdot\vect{m}\> dxdy \>\vect{z} \\
	\left[\mathcal{D}\right]&=	
	\begin{bmatrix}
		\mathcal{D}_{\rm{xx}} & \mathcal{D}_{\rm{xy}} \\
		\mathcal{D}_{\rm{yx}} & \mathcal{D}_{\rm{yy}}
	\end{bmatrix},\mathcal{D}_{ij}=\frac{M_{\rm{S}}t_{\rm{FM}}}{\gamma}\iint\interpar{\frac{\partial \vect{m}}{\partial i}\cdot\frac{\partial \vect{m}}{\partial j}}\> dxdy \\
F_{\rm{x,y}}&=\frac{\mu_0M_{\rm{S}}t_{\rm{FM}}}{\gamma}\iint\interpar{\vect{m}\times\vect{\Gamma}}\cdot\frac{\partial \vect{m}}{\partial x,y}\> dxdy
\end{aligned}
\end{equation}
where $\vect{G}$ is the gyrovector, $\gamma$ is the gyromagnetic ratio, $\alpha$ is the Gilbert damping parameter, $\left[\mathcal{D}\right]$ the dissipation matrix, $\vect{v}$ the skyrmion velocity and $\vect{F}$ the force exerted on the skyrmion magnetization by the torque $\vect{\Gamma}$. In the present example, we model the injection of a vertical spin current polarized along y, which exerts a torque $\vect{\Gamma}_{\rm{SOT}}$. For simplicity, we only consider a damping-like component, as the field-like component may deform the skyrmion but shall not move it, in first approximation. This would correspond, for example, to the case of a spin Hall effect due to a charge current flowing along $x$ in the multilayer. As we consider an axisymmetric skyrmion with $m_{\rm{z}}(0)=-\vect{z}$ and $m_{\rm{z}}(\infty)=+\vect{z}$ (see Fig.\ \ref{fig:Scheme}), it results that off-diagonal terms in $\left[\mathcal{D}\right]$ are zero and
\begin{equation}
\begin{aligned}
	\vect{G}=&4\pi \frac{M_{\rm{S}}t_{\rm{FM}}}{\gamma} \\
	\mathcal{D}_{\rm{xx,yy}}=&\pi\frac{M_{\rm{S}}t_{\rm{FM}}}{\gamma}a \\
	F_{\rm{x,y}}=&\mp\theta_{\rm{SH}}J\frac{\pi\hbar}{2e}b_{\rm{x,y}}
\end{aligned}
\end{equation}
where $a$ is a dimensionless coefficient related to the magnetic structure of the skyrmion; $J$ is the in-plane charge current density in the multilayer and $\theta_{\rm{SH}}$ is the effective spin Hall angle; $b_{\rm{x}}$ and $b_{\rm{y}}$ are homogeneous to different characteristic sizes of the skyrmion, related to their geometry and magnetic structures, defined as 
\begin{equation}
\begin{aligned}
\label{eq:Fxyfactors}
	a=&\int_{r=0}^{\infty} \frac{1-m_{\rm{z}}^2}{r}+r\sum_{i=x,\varphi,z}\interpar{\frac{\partial m_{\rm{i}}}{\partial r}}^2 dr \\
	b_{\rm{x}}=&\int_{r=0}^{\infty}r\interpar{\frac{\partial m_{\rm{r}}}{\partial r}m_{\rm{z}}-\frac{\partial m_{\rm{z}}}{\partial r}m_{\rm{r}}}+m_{\rm{r}}m_{\rm{z}}\>dr \\
	b_{\rm{y}}=&\int_{r=0}^{\infty}r\interpar{\frac{\partial m_{\rm{\varphi}}}{\partial r}m_{\rm{z}}-\frac{\partial m_{\rm{z}}}{\partial r}m_{\rm{\varphi}}}+m_{\rm{\varphi}}m_{\rm{z}}\>dr
\end{aligned}
\end{equation}
where details of the derivation are given in Appendix \ref{app:Thielevectors}. We note that the formalism provided by the Thiele equation is not able to treat the effect of deformations of the skyrmions. As a consequence, the following conclusions are valid for small, compact skyrmions at low current densities, but may be altered by for more deformable, plateau skyrmions (with a uniform magnetization core of finite size) at higher current densities.

\begin{figure*}
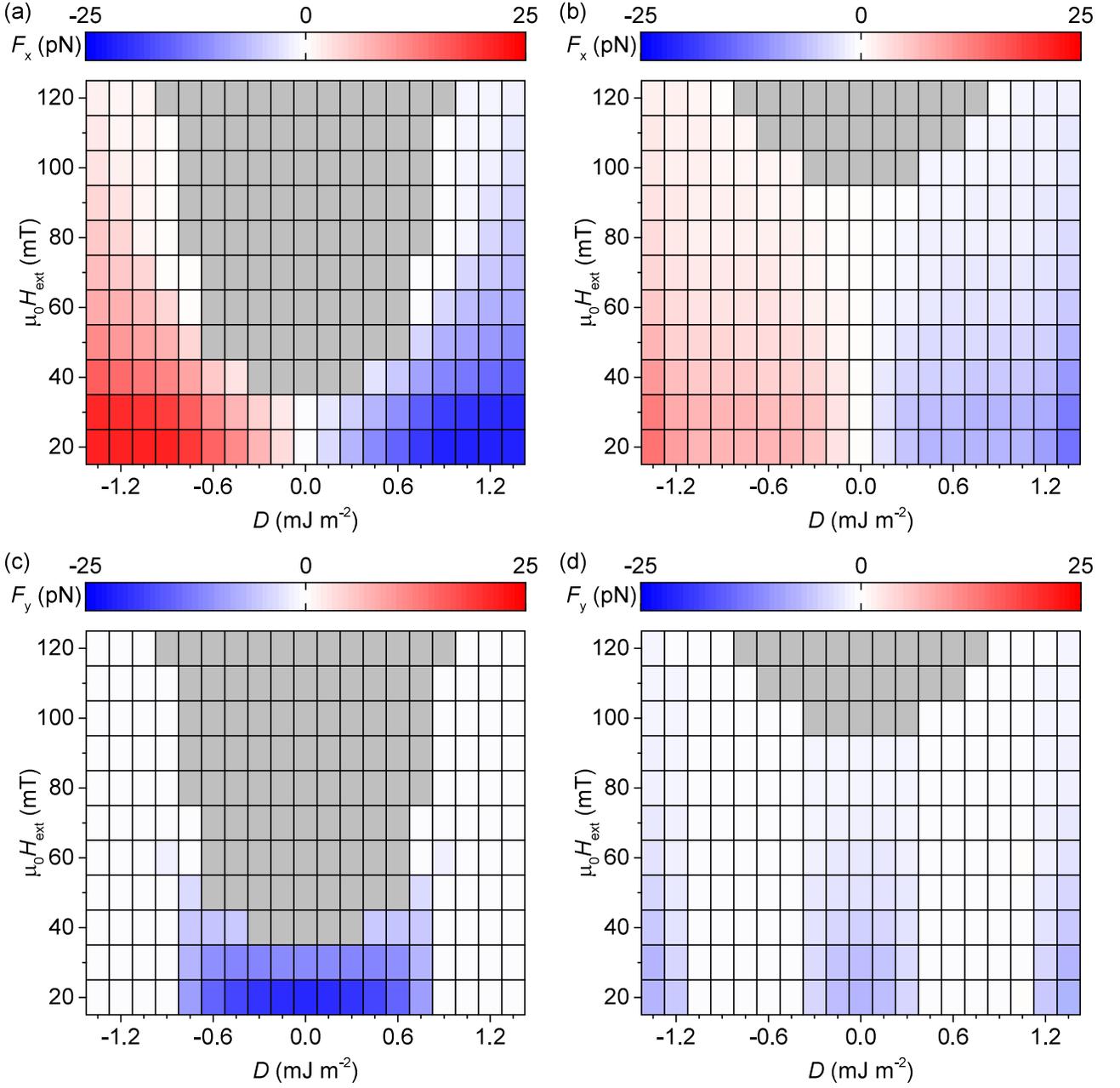

\includegraphics[width=3.375in, trim= 0cm 0cm 0cm 0cm]{{{Sk_Fx_BD_t=1.4e-3_1D}}}
\includegraphics[width=3.375in, trim= 0cm 0cm 0cm 0cm]{{{Sk_Fx_BD_t=1.4e-3_Full}}}
\includegraphics[width=3.375in, trim= 0cm 0cm 0cm 0cm]{{{Sk_Fy_BD_t=1.4e-3_1D}}}
\includegraphics[width=3.375in, trim= 0cm 0cm 0cm 0cm]{{{Sk_Fy_BD_t=1.4e-3_Full}}}
\caption{Forces acting on the skyrmion due to the spin-orbit torques as a function of DMI parameter and external field, along $\vect{x}$ for (a) a model with uniform magnetization through the thickness and (b) a layer-by-layer description, along $\vect{y}$ for (c) a model with uniform magnetization through the thickness and (d) a layer-by-layer description. The sign and magnitude of the force is given by the color scale. The gray color signifies that the skyrmion collapses below the cell size for these parameters.}
\label{fig:BDForces}
\end{figure*}

As can be seen from \eqref{eq:Fxyfactors}, the direction of the force driving the current-induced motion is directly set by the texture and chirality of the skyrmion: a N\'eel skyrmion is driven by a force along the $\vect{x}$ direction while a Bloch skyrmion is driven by a force along the $\vect{y}$ direction, with the sign of the force being determined by the chirality. As a consequence, the layer-dependent chirality obtained in the layer-by-layer approach is crucial to model correctly the current-induced motion of hybrid chiral skyrmions.

We present in Fig.\ \ref{fig:BDForces} the results of the Thiele modeling for a system identical to the one of the previous section ($L=3$). We consider a multilayer geometry in which a similar spin current of magnitude given by $\theta_{\rm{SH}}J =$ \SI{2e10}{\ampere\per\meter\squared} is injected into each ferromagnetic layer, which would correspond to the case of having an identical heavy-metal layer adjacent to each ferromagnetic layer. The obtained effective forces $F_{\rm{x}}$ [Figs.\ \ref{fig:BDForces}(a) and \ref{fig:BDForces}(b)] and $F_{\rm{y}}$ [Figs.\ \ref{fig:BDForces}(c) and \ref{fig:BDForces}(d)] due to current-induced torques are the direct consequence of the average of the chirality in all layers shown in Figs.\ \ref{fig:BDProfiles}(c) and \ref{fig:BDProfiles}(d). We thus compare again a model with uniform magnetization through the thickness [Figs.\ \ref{fig:BDForces}(a) and \ref{fig:BDForces}(c)] and the layer-by-layer model [Figs.\ \ref{fig:BDForces}(b) and \ref{fig:BDForces}(d)]. We find, in the case of layer-by-layer modeling, that the driving force is maximized for a uniform chirality, while it is strongly reduced for a hybrid chirality, due to N\'eel skyrmions with opposite chiralities. Indeed, in the latter case top and bottom layer skyrmions, showing opposite chiralities, are driven in opposite directions. This is the main reason for the large difference in the magnitude of the driving forces between the two models. Because the uniform model fails to predict and describe the hybrid chirality, and hence neglects the competing current-induced forces arising in top and bottom layers, it largely overestimates the effect of the spin currents on the global structure for all values of $D$ that are not large enough to ensure a unique chirality in the whole stack of layers. 

\section{Discussion}

\begin{figure}
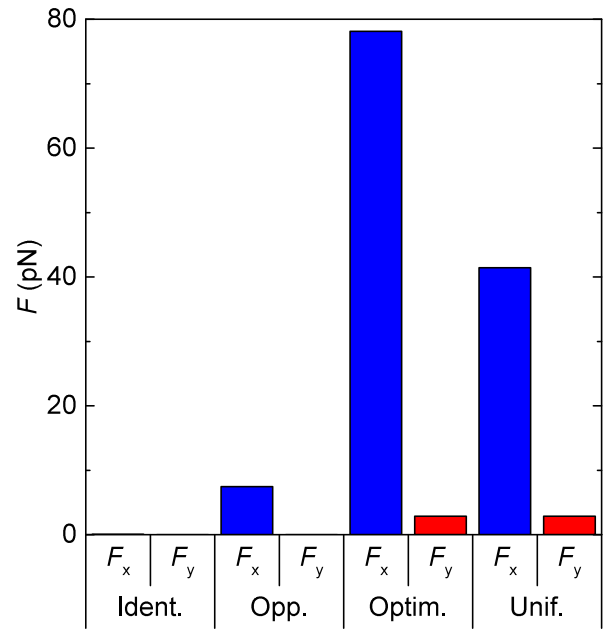

\includegraphics[width=3.375in, trim= 0cm 0cm 0cm 0cm]{{{Sk_Fxy_geom_N=20}}}
\caption{Forces $F_{\rm{x}}$ and $F_{\rm{y}}$ acting on the multilayered hybrid chiral skyrmion due to the spin-orbit torques in different geometries of spin-injection, for the case of $L=20$ and $\mu_0H_{\rm{ext}} =$ \SI{200}{\milli\tesla}.}
\label{fig:L20Forces}
\end{figure}

We have recently shown the importance of taking into account the spin torque symmetries with respect to the effective skyrmion chiralities in multilayers in order to achieve a fast propagation of the skyrmions\cite{Legrand2018}. Beyond the case of this simple, pedagogical example of a uniform current injection geometry with $L=3$ (see Fig.\ \ref{fig:BDForces}, our model can be used to investigate any type of spin-current injection scheme in multilayers, allowing us to predict the dynamics of skyrmions and discriminate the interesting configurations among a very large choice of possible experimental multilayered structures. For example, we consider here different spin-current injection geometries in the case of a complex, hybrid skyrmion as found for $L=20$ and $\mu_0H_{\rm{ext}} =$ \SI{200}{\milli\tesla} (Fig.\ \ref{fig:L20profiles}). The spin current magnitude is again $\abs{\theta_{\rm{SH}}J} =$ \SI{2e10}{\ampere\per\meter\squared} in all cases. The obtained forces are shown in Fig.\ \ref{fig:L20Forces}.

In the first geometry (labelled as ``Ident.\,'' in Fig.\ \ref{fig:L20Forces}), the spin-currents generate an identical spin-accumulation in the topmost and bottommost layers only. This situation corresponds, for example, to spin-orbit torques existing in multilayers being enclosed between two heavy-metal layers of opposite spin Hall angles ({\it e.g.}, Pt and Ta), with negligible spin-torques generated inside the multilayer. In this geometry, due to the hybrid chirality, the forces exerted on top and bottom parts of the skyrmion cancel each other so that the total force drops to zero. In the second geometry (labelled as ``Opp.\,'' in Fig.\ \ref{fig:L20Forces}), opposite spin-injections occur at the top and bottom layers of the multilayer. This situation corresponds, for example, to spin-orbit torques existing in multilayers being enclosed between two heavy-metal layers of identical spin Hall angles. In this geometry, due to the hybrid chirality, the forces exerted on top and bottom parts of the skyrmion add up so that the resulting force becomes significant. In the third geometry (labelled as ``Optim.\,'' in Fig.\ \ref{fig:L20Forces}), the sign of the spin-injection is designed to match the effective chirality in each layer as it is shown in Fig.\ \ref{fig:L20profiles}. This can be achieved by inserting a heavy-metal of the appropriate sign of spin Hall angle adjacent to each ferromagnetic layer. In this case, the torques are optimized and result in an about tenfold increase of the skyrmion driving forces as all layers (except $l_{15}$ hosting a Bloch skyrmion) now contribute to the total force. In the last studied geometry (labelled as ``Unif.\,'' in Fig.\ \ref{fig:L20Forces}), the spin-injection is identical in all layers. This is simpler to achieve as it consists in repeating the same heavy-metal adjacent to each ferromagnetic layer. The total force resulting of all layer-dependent forces thus reflects the balance between layers hosting N\'eel skyrmions with $\phi=0$ and $\phi=\pi$. In this case, the driving force is reduced by half as compared to the previous case, as only ten layers contribute to the total force, while the forces exerted in the five bottom-most and five top-most layers, hosting skyrmions of opposite chiralities, cancel each other. In the last two cases, a small force directed along $y$ results from the Bloch part of the multilayered skyrmion, located in one of the intermediate layers. The forces are much weaker for the geometry ``Opp.'' than for ``Optim.'', because for the first the spin currents act on the topmost and bottommost layers only, rather than on all 20 layers. If multiplied by this factor of ten, $F_{\rm{x}}$ for ``Opp'' is about $F_{\rm{x}}$ for the ``Optim.'' geometry. These results, using the Thiele formalism, are in excellent agreement with our recent micromagnetic simulations\cite{Legrand2018}.

K.\-W.\ Kim\cite{Kim2018c} et al have very recently suggested that the trajectory of skyrmions relative to the applied current could be engineered in hybrid-DMI systems, showing DMI of both bulk and interfacial origin, thus favoring neither pure Bloch nor pure N\'eel skyrmions. Most notably, it would be possible in these systems to cancel out their skyrmion Hall effect, that is, the side motion of the skyrmions due to the gyrotropic term in their dynamics. Here, our model suggests that even in the case of a uniform, purely interfacial DMI, the presence of hybrid chirality can be exploited to control the skyrmion Hall angle of multilayered skyrmions. To support this statement, we use the forces as determined from the Thiele equation in the previous section, displayed in Fig.\ \ref{fig:BDForces}, in order to find the skyrmion velocity and skyrmion Hall angle in the simple case of $L=3$ with an uniform spin-injection in all layers.

\begin{figure}
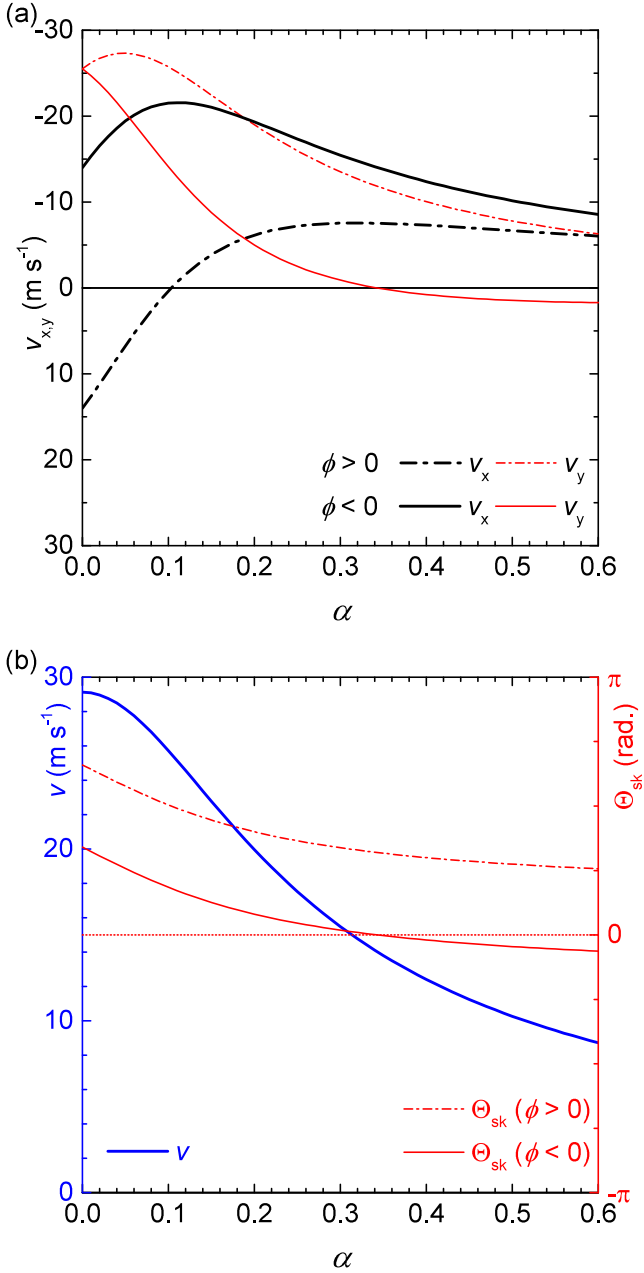

\includegraphics[width=3.375in, trim= 0cm 0cm 0cm 0cm]{{{Sk_vxy_alpha_N=3}}}
\includegraphics[width=3.375in, trim= 0cm 0cm 0cm 0cm]{{{Sk_vth_alpha_N=3}}}
\caption{(a) Longitudinal $v_{\rm{x}}$ (black, thick lines) and transverse $v_{\rm{y}}$ (red, thin lines) velocities as a function of the Gilbert damping $\alpha$, for a hybrid chiral skyrmion with $\phi>0$ (dashed-dotted lines) and $\phi<0$ (solid lines). (b) Skyrmion velocity $v$ (left scale, thick blue line), and skyrmion Hall angle $\Theta_{\rm{sk}}$ (right scale, red thin lines) for a hybrid chiral skyrmion with $\phi>0$ (dashed-dotted lines) and $\phi<0$ (solid lines). }
\label{fig:L3velocities}
\end{figure}

First, we display in Fig.\ \ref{fig:L3velocities}(a) both $v_{\rm{x}}$ (black, thick lines) and $v_{\rm{y}}$ (red, thin lines) as a function of Gilbert damping $\alpha$, obtained by solving \eqref{eq:ThieleEq} for $D =$ \SI{1.35}{\milli\joule\per\meter\squared} and $\mu_0H_{\rm{ext}} =$ \SI{50}{\milli\tesla} [see the corresponding skyrmion configuration in Figs.\ \ref{fig:BDProfiles}(b) and \ref{fig:BDProfiles}(d),  and forces in Figs.\ \ref{fig:BDForces}(b) and \ref{fig:BDForces}(d)]. We consider two cases depending on the chiral configuration of the top layer. Indeed two configurations are actually degenerate in energy, for positive values (dashed-dotted lines) and negative values (solid lines) of the angle $\phi(r)$, which corresponds to the Bloch component of the skyrmion having one chirality or the other (still, the chirality of the N\'eel component is fixed by the DMI). The values of the forces $F_{\rm{x}}$ and $F_{\rm{y}}$ are constant and determined by the texture of the skyrmion. In addition, depending on whether $\phi>0$ or $\phi<0$, the transverse force $F_{\rm{y}}$ changes sign. Overall, in combination with the effect of the gyrotropic term that rotates the direction of motion, both $v_{\rm{x}}$ and $v_{\rm{y}}$ evolve significantly with $\alpha$. Actually, the motion is deflected by an angle in between $90^{\circ}$ (in the limit of $\alpha\rightarrow0$) and $0^{\circ}$ (in the limit of $\alpha\rightarrow\infty$). As a result, in one case $v_{\rm{y}}$ crosses zero. This appears clearly in Fig.\ \ref{fig:L3velocities}(b), which displays the skyrmion velocity $v$ (left axis, in blue) and the skyrmion Hall angle $\Theta_{\rm{sk}}$ (right axis, in red) as a function of $\alpha$. In the case of $\phi<0$, for $\alpha\approx0.35$, the skyrmion Hall effect is canceled while a significant velocity is still achieved. Considering the consequence of the hybrid chirality of skyrmions on their dynamics thus provide a means of canceling the undesired, transverse motion. By combining this approach with the engineering of the spin-injection geometry, a high degree of control on the trajectories can be achieved.

We believe that our model constitutes a useful tool in order to further reduce the size of magnetic skyrmions, considering simultaneously all aspects among detailed layer-by-layer magnetization profile, stability, and optimal spin current injection for motion. As can be seen from the difference between Figs.\ \ref{fig:BDProfiles}(a) and \ref{fig:BDProfiles}(b), the predicted sizes and field stability of the skyrmions differ much depending on whether the multilayered nature of the skyrmions is considered. From an application point of view, our model allows to accurately predict the expected skyrmion properties for different multilayer compositions. It will thus allow to optimize the material design of skyrmion multilayers, which opens the way for the development of various skyrmion based devices, going from skyrmion racetrack memories\cite{Fert2017} to neuro-inspired component architectures\cite{Prychynenko2018,Zazvorka2018arXiv}.

\section{Summary}

In conclusion, we propose a layer-by-layer model, which is able to predict accurately the actual profile of multilayered magnetic skyrmions and to describe the effects of interlayer dipolar interactions on the skyrmion size, stability, and chirality. Notably, this model is extremely suitable to study hybrid chirality through the magnetic layers of multilayered systems. Using the Thiele formalism, we have shown how this model can be used to predict the skyrmion velocity and skyrmion Hall angle under current-driven motion for any kind of spin-current injection scheme.

Even if we have focused here on the case of multilayers made of repetitions of a unique base of identical layers (identical ferromagnetic element(s), layer thicknesses and spacer thicknesses and materials), we note that it can also be used to model the case of layer-by-layer varying magnetic parameters inside a stacking of different ferromagnetic layers. By including direct exchange between contiguous layers, and other forms of DMI, this model could also be easily adapted to the description of thick skyrmions in chiral magnets\cite{Yu2010a,Montoya2017}.

The present model can also be easily further refined by adding any other magnetic interaction, which opens the perspective of describing accurately more specific systems. For example, considering interlayer couplings mediated by electrons would allow to treat the case of thin metallic spacers with RKKY interactions and thus the cases of coupled bilayers or antiferromagnetically coupled layers\cite{Zhang2016,Hrabec2017a}. 

Obtaining the accurate layer-by-layer profile of the multilayered skyrmions shall also have implications, for example, on their electrical detection using their magnetic texture\cite{Hanneken2015, Maccariello2018} or even on their thermodynamical stability\cite{Rohart2016,vonMalottki2017,Bessarab2018}.

\begin{acknowledgments}
This work was partially supported by the Agence Nationale de la Recherche, France, under grant agreement No. ANR-17-CE24-0025 (TOPSKY), by the Horizon2020 Framework Programme of the European Commission, under grant agreement No. 665095 (MAGicSky) and the DARPA TEE program through grant MIPR\# HR0011831554.
\end{acknowledgments}

\appendix

\section{Solution for the magnetic potential}
\label{app:Solution_poisson}

Solution for $\bar{\psi_{\rm{f}}}$. We consider as a general solution
\begin{equation}
\bar{\psi_{\rm{f}}}=\left\{
	\begin{aligned}
	&B_1\cosh{\interpar{kz}}+B_2\sinh{\interpar{kz}}+B_0, \text{inside} \\
	&A_1\exp{\interpar{kz}}+A_2\exp{\interpar{-kz}}+A_0, z>t/2
	\end{aligned}\right.
\end{equation}
for the inside and outside parts, respectively, with $A_{0,1,2},B_{0,1,2}$ real constants. Using (i), we find that $B_0=-\bar{f}(k)/k^2$ and $A_0=0$. Condition (ii) yields $A_1=0$. Using the symmetry with respect to z of volume charges, we get that $\bar{\psi_{\rm{f}}}=A_2\exp{\interpar{kz}}, z<-t/2$, and $B_2=0$. Combining (iii) and (iv) at $z=t/2$ we find 
\begin{equation}
\left\{
	\begin{aligned}
	&B_1\cosh{\interpar{kt/2}}-\bar{f}(k)/k^2=A_2\exp{\interpar{-kt/2}} \\
	&kB_1\sinh{\interpar{kt/2}}=-kA_2\exp{\interpar{-kt/2}}
	\end{aligned}
\right.
\end{equation}
which finally gives $A_2$ and $B_1$.

Solution for $\bar{\psi_{\rm{h}}}$. We consider as a general solution
\begin{equation}
\bar{\psi_{\rm{h}}}=\left\{
	\begin{aligned}
	&B_1\cosh{\interpar{kz}}+B_2\sinh{\interpar{kz}}+B_0, \text{ inside} \\
	&A_1\exp{\interpar{kz}}+A_2\exp{\interpar{-kz}}+A_0, z>t/2
	\end{aligned}\right.
\end{equation}
for the inside and outside parts, respectively, with $A_{0,1,2},B_{0,1,2}$ real constants. Using (i), we find that $B_0=0$ and $A_0=0$. Condition (ii) yields $A_1=0$. Using the antisymmetry with respect to z of surface charges, we get that $\bar{\psi_{\rm{h}}}=-A_2\exp{\interpar{kz}}, z<-t/2$, and $B_1=0$. Combining (iii) and (iv) at $z=t/2$ we find 
\begin{equation}
\left\{
	\begin{aligned}
	&B_2\sinh{\interpar{kt/2}}=A_2\exp{\interpar{-kt/2}} \\
	&kB_2\cosh{\interpar{kt/2}}+\bar{h}(k)=-kA_2\exp{\interpar{-kt/2}}
	\end{aligned}
\right.
\end{equation}
which finally gives $A_2$ and $B_2$.

\section{Considerations for numerical implementation}
\label{app:Condnum}

A first difficulty is the correct and efficient implementation of the Hankel transform. Indeed, the Bessel functions are slowly decaying and the use of a large set of values of $k$ is required for a correct reconstruction of the functions. In case of undersampling, the Hankel transform is no longer involutive and $\bar{\bar{\psi}}\neq\psi$ which conveys wrong results. For the implementation of the Hankel transform, it is a good idea to choose sampling $r$ points related to the position of the zeros of the Bessel functions\cite{Johnson1987}. This way, it is possible to obtain very accurate transforms even for a number of $k$ points equal to the number of $r$ points and relatively steep functions. Such method has been tested and validated. 

Whether a regular spacing or a specific choice of $r$ points is chosen, we note that the use of functions for which $\int_0^\infty \abs{g(x)}x^{1/2}dx$ converges is required. For isolated skyrmions, which have a limited radius and are stabilized in an otherwise uniformly magnetized layer, for example along $+\vect{z}$, this is the case of $f$, but not $h$. For this reason, we define $\tilde{h}=M_{\rm{S}}(1-\vect{m}\cdot\vect{z})$, the surface charges associated to the difference $\vect{m}(r)-\vect{m_{\rm{unif}}}(r)$ with $\vect{m_{\rm{unif}}}(r)=\vect{z}$ the uniform magnetization. This new $\tilde{h}$ function equals zero outside the skyrmion and has a properly defined transform. As the dipolar field associated to a source $\vect{m_{\rm{unif}}}(r)$ is confined inside the layer and is equal to $-M_{\rm{S}}\vect{z}$, by summing the two source terms $\vect{m}(r)-\vect{m_{\rm{unif}}}$ and $\vect{m_{\rm{unif}}}(r)$ we find that 
\begin{equation}
\vect{H_{\rm{dip}}} =\left\{
\begin{aligned}
&\vect{\tilde{H}_{\rm{dip}}}-M_{\rm{S}}\vect{z},& -t/2 \leq z \leq t/2 \\
&\vect{\tilde{H}_{\rm{dip}}},& \abs{z}>t/2
\end{aligned}\right.
\end{equation}
with $\vect{\tilde{H}_{\rm{dip}}}$ the partial field obtained by replacing $h$ by $\tilde{h}$.

The energy of the system includes the dipolar term, which is a long-range interaction and has an infinite extension. For an isolated skyrmion, a part of the stabilizing dipolar energy thus comes from the interaction with magnetic moments located towards infinity. However, for numerical implementation we have to consider a finite-size simulation space. We suppose $N$ points $r_1,r_2,\dots,r_N>0$, and as boundary conditions, that $\vect{m}=-\vect{z}$ at $r=0$ and $\vect{m}=\vect{z}$ for $r>r_N$. With our definition of $E_{\rm{sk}}$, the integrand of \eqref{eq:Esk} is zero for $r>r_N$ with $\vect{m}=\vect{z}$, except for the dipolar energy term, due to the long-range nature of dipolar interactions. In order to be able to find the energy of the system, we then have to transform this expression in order to find an integrand which becomes zero for $r>r_N$ . We separate the system into two parts, interior with variable $\vect{m}$ for $r\leq r_N$ and exterior with fixed $\vect{m}$ for $r>r_N$. We can write 
\begin{equation}
\vect{E_{\rm{dip}}} = \interpar{\int_{int} + \int_{ext}} \frac{-\mu_0M_{\rm{S}}}{2}\interpar{\vect{H_{\rm{var}}}+\vect{H_{\rm{fix}}}}\cdot\vect{m}
\end{equation}
with $\vect{H_{\rm{var}}}$ the fields generated by the interior part and $\vect{H_{\rm{fix}}}$ the fields generated by the exterior part. In the exterior part, both $\vect{m}$ and $\vect{H_{\rm{fix}}}$ are fixed, so that their scalar product is constant and can be dropped. We then keep 
\begin{equation}
\vect{E_{\rm{dip}}} = \frac{-\mu_0M_{\rm{S}}}{2} \left[\int_{int} \interpar{\vect{H_{\rm{var}}}+\vect{H_{\rm{fix}}}}\cdot\vect{m} +\int_{ext} \vect{H_{\rm{var}}}\cdot\vect{m} \right]
\end{equation}
and, as it is an interaction between two sub-systems, we get that
\begin{equation}
\int_{ext} \vect{H_{\rm{var}}}\cdot\vect{m} =\int_{int} \vect{H_{\rm{fix}}}\cdot\vect{m}
\end{equation}
so that we can write
\begin{equation}
\vect{E_{\rm{dip}}}=\frac{-\mu_0M_{\rm{S}}}{2} \int_{int} \vect{H_{\rm{dip}}}\cdot\vect{m} + \vect{H_{\rm{fix}}}\cdot\vect{m}
\end{equation}
In this formula, $\vect{H_{\rm{fix}}}$ is then the field generated by a saturated outer part in the interior part. It can be obtained as $-M_{\rm{S}}\vect{z}-\vect{H_{\rm{var}}}(h=1)$, where $\vect{H_{\rm{var}}}$ is found from a saturated inner part ($h=1$ for $r<r_N$). Finally, \eqref{eq:Eskred} is redefined as an integral between 0 and $r_N/\sqrt{A/K}$ with all terms unchanged except the dipolar part, for which the integrand becomes
\begin{equation}
\frac{-\mu_0M_{\rm{S}}}{2K_{\rm{eff}}} \interpar{M_{\rm{S}}+\vect{H_{\rm{dip}}}\cdot\vect{m} + \vect{H_{\rm{fix}}}\cdot\interpar{\vect{m}-\vect{z}}}
\end{equation}

We note that the forward and backward Hankel transforms, plus mathematical operations required in Eqs.\ \eqref{eq:psibarsol},\eqref{eq:Hdip} to find the field of each layer acting on each layer, at each evolution step are extremely costly as compared to what is required for the four other field terms. As the Hankel transforms are matrices multiplications, they can be factorized with the formulae of \eqref{eq:psibarsol}. Moreover, the dipolar field can be calculated by performing only once the determination of the factors inside $\psi$ and deducing $\vect{H_{\rm{dip}}}$ for all layers in a last step. Very importantly, there is actually no need to repeat these operations at each evolution step. Because magnetostatic problems can be solved by superposition of simpler problems, as we have done with $f$ and $h$, we can also do it for the source terms related to $\vect{m}(r)$ at each $r$. Given a sampling of $r$ points $r_1,r_2,\dots,r_N$, we will precompute four dipolar field kernels $K_{\rm{h,f}}^{r,z}$ for the dipolar fields due to $f$ and $h$ distributions, along $r$ and $z$. They are $L\times L \times N \times N$ arrays where $K_{\rm{h,f}}^{r,z}(j,i,r_j,r_i)$ store the magnetic fields in layer $j$ at point $r_j$ due to a unit (surface or volume) charge located in layer $i$ at point $r_i$. By superposition of each elementary source, we find that the fields in each layer $j$ are
\begin{equation}
\begin{aligned}
	\begin{bmatrix}
		H_{\rm{dip}}^{r,z}(r_1) \\ 
		H_{\rm{dip}}^{r,z}(r_2) \\ 
		\vdots \\
		H_{\rm{dip}}^{r,z}(r_N) \\ 
	\end{bmatrix}
	=&\sum_{i=1}^L K_f^{r,z}(j,i,r_1\cdots r_N,r_1\cdots r_N)
	\begin{bmatrix}
		f_i(r_1) \\ 
		f_i(r_2) \\ 
		\vdots \\
		f_i(r_N) \\ 
	\end{bmatrix} \\
	+&\sum_{i=1}^L K_h^{r,z}(j,i,r_1\cdots r_N,r_1\cdots r_N)
	\begin{bmatrix}
		h_i(r_1) \\ 
		h_i(r_2) \\ 
		\vdots \\
		h_i(r_N) \\ 
	\end{bmatrix} \\
\end{aligned}
\end{equation}
which gives self-interacting or interaction between two layers $\vect{H_{\rm{dip}}}$ in four matrix multiplications. Because there is no cost of having a fine sampling of $k$ points once the kernels are computed, we make the choice of using a regular spacing of r points in order to get easier computation of other fields, and comparison with Mumax$^3$, at the cost of some more initialization time.

Finally, we have chosen above a Cartesian description of $\vect{m}=\interpar{m_{\rm{r}},m_{\rm{\varphi}},m_{\rm{z}}}$. As the magnetization $\vect{m}$ is normalized and evolves on the unit sphere, a description in spherical coordinates $\vect{m}=\interpar{\sin(\theta)\cos(\phi),\sin(\theta)\sin(\phi),\cos(\theta)}$, with $\theta$ polar and $\phi$ azimuthal angles, can be convenient and is often used for describing the profile of magnetic skyrmions. We provide in Appendix \ref{app:Heffspher} the expression of all fields and evolution step within spherical coordinates. Using it for the determination of the profiles is nevertheless not extremely suitable, as there is a definition issue for the azimuthal angle $\phi$ around $\vect{m}=\pm\vect{z}$ which degrades the convergence of the solver, and an additional computational complexity of involving multiple $\cos$ and $\sin$ functions. Because of these issues, we always use, in the present work, Cartesian coordinates for computing evolution steps (and maintain a unit magnetization everywhere), even if it may be convenient to display the final results with polar representation.

\section{Fields with $\vect{m}$ in spherical coordinates}
\label{app:Heffspher}

Effective fields associated to exchange, DMI and anisotropy
\begin{equation}
H_{\rm{A}}=\frac{2A}{\mu_0M_{\rm{S}}}\begin{bmatrix}
	\frac{d^2\theta}{dr^2}\cos{\theta}\cos{\phi}-\frac{d^2\phi}{dr^2}\sin{\theta}\sin{\phi}\\
	-\interpar{\frac{d\theta}{dr}}^2\sin{\theta}\cos{\phi}-\interpar{\frac{d\phi}{dr}}^2\sin{\theta}\cos{\phi}\\
	-2\frac{d\theta}{dr}\frac{d\phi}{dr}\cos{\theta}\sin{\phi}+\frac{1}{r}\frac{d\theta}{dr}\cos{\theta}\cos{\phi}\\
	-\frac{1}{r}\frac{d\phi}{dr}\sin{\theta}\sin{\phi}-\frac{\sin{\theta}\cos{\phi}}{r^2}\\
	\\
	\frac{d^2\theta}{dr^2}\cos{\theta}\sin{\phi}+\frac{d^2\phi}{dr^2}\sin{\theta}\cos{\phi}\\
	-\interpar{\frac{d\theta}{dr}}^2\sin{\theta}\sin{\phi}-\interpar{\frac{d\phi}{dr}}^2\sin{\theta}\sin{\phi}\\
	+2\frac{d\theta}{dr}\frac{d\phi}{dr}\cos{\theta}\cos{\phi}+\frac{1}{r}\frac{d\theta}{dr}\cos{\theta}\sin{\phi}\\
	+\frac{1}{r}\frac{d\phi}{dr}\sin{\theta}\cos{\phi}-\frac{\sin{\theta}\sin{\phi}}{r^2} \\
	\\
	-\frac{d^2\theta}{dr^2}\sin{\theta}-\interpar{\frac{d\theta}{dr}}^2\cos{\theta}-\frac{1}{r}\frac{d\theta}{dr}\sin{\theta}	
\end{bmatrix}
\end{equation}

\begin{equation}
H_{\rm{D}}=\frac{-2D}{\mu_0M_{\rm{S}}}\begin{bmatrix}
	\sin{\theta}\frac{d\theta}{dr} \\
	0\\
	\frac{d\theta}{dr}\cos{\theta}\cos{\phi}-\frac{d\phi}{dr}\sin{\theta}\sin{\phi}+\frac{\sin{\theta}\cos{\phi}}{r}
\end{bmatrix}
\end{equation}

\begin{equation}
H_{\rm{K}}=\frac{2K_{\rm{u}}}{\mu_0M_{\rm{S}}}\cos{\theta}\vect{z}
\end{equation}
Integrands of \eqref{eq:Esk} associated to exchange, DMI and anisotropy energies
\begin{equation}
\begin{aligned}
	E_{\rm{A}}&=A\left[\interpar{\frac{d\theta}{dr}}^2+\interpar{\frac{d\theta}{dr}}^2\sin{\theta}+\frac{\sin^2{\theta}}{r^2}\right] \\
	E_{\rm{D}}&=D\left[\frac{d\theta}{dr}\cos{\phi}-\frac{d\phi}{dr}\cos{\theta}\sin{\theta}\sin{\phi}+\frac{\cos{\theta}\sin{\theta}\cos{\phi}}{r}\right] \\
	E_{\rm{K}}&=K_{\rm{u}}\cos^2{\theta}
\end{aligned}
\end{equation}
Magnetic charges for the determination of the dipolar field
\begin{equation}
\begin{aligned}
f&=M_{\rm{S}}\interpar{\frac{d\theta}{dr}\cos{\theta}\cos{\phi}-\frac{d\phi}{dr}\sin{\theta}\sin{\phi}+\frac{\sin{\theta}\cos{\phi}}{r}} \\
h&=-M_{\rm{S}}\cos{\theta}
\end{aligned}
\end{equation}
Evolution steps for $(\theta,\phi)$
\begin{equation}
\begin{aligned}
\delta_{\theta}&= \lambda\interpar{H_{\rm{r}}\cos{\theta}\cos{\phi}+H_{\rm{\varphi}}\cos{\theta}\sin{\phi}+H_{\rm{z}}\sin{\theta}} \\
\delta_{\phi}&= \frac{\lambda}{\sin{\theta}}\interpar{H_{\rm{\varphi}}\cos{\phi}-H_{\rm{r}}\sin{\phi}}
\end{aligned}
\end{equation}

\section{Energy of the domain profile in Cartesian coordinates}
\label{app:Eskcart}

We define here the Cartesian profile of a double domain-wall with $m_{\rm{x}}(\pm x)=\pm m_{\rm{r}}(r)$, $m_{\rm{y}}(\pm x)=m_{\rm{\varphi}}(r)$ and $m_{\rm{z}}(\pm x)= m_{\rm{z}}(r)$. The energies associated to symmetric exchange interaction, DMI, anisotropy, Zeeman fields are straightforward, while the dipolar field is found as in section \ref{subsec:dipolarsol}, using the usual Fourier transform instead of the Hankel transform, with $h(x)=-M_{\rm{S}}m_{\rm{z}}(x)$ and $f(x)=M_{\rm{S}}\partial m_{\rm{x}}(x)/\partial x$. In the k-space, all expressions are unchanged. We get, now using $\chi=x/\sqrt{A/K}$,
\begin{widetext}
\begin{multline}
E_{\rm{st}}=t\sqrt{AK_{\rm{eff}}}\int_{-\infty}^{\infty} \left\{\interpar{\frac{dm_{\rm{x}}}{d\chi}}^2+\interpar{\frac{dm_{\rm{y}}}{d\chi}}^2+ \interpar{\frac{dm_{\rm{z}}}{d\chi}}^2 +\frac{4D}{\pi D_{\rm{c}}}\interpar{m_{\rm{z}}\frac{dm_{\rm{x}}}{d\chi}-m_{\rm{x}}\frac{dm_{\rm{z}}}{d\chi}}\right. \\
+\left.\frac{K_{\rm{u}}}{K_{\rm{eff}}}\interpar{1-m_{\rm{z}}^2} + \frac{\mu_0H_{\rm{ext}}M_{\rm{S}}}{K_{\rm{eff}}}\interpar{1-m_{\rm{z}}} - \frac{\mu_0M_{\rm{S}}^2}{2K_{\rm{eff}}}\interpar{1+\frac{\vect{H_{\rm{dip}}}}{M_{\rm{S}}}\cdot\vect{m}}\right\} d\chi
\end{multline}
\end{widetext}

\section{Derivation of the vectors composing the Thiele equation}
\label{app:Thielevectors}
For a magnetization vector $\vect{m}(x,y)$ inside the structure we define $r=\sqrt{x^2+y^2}$ and $\varphi$ so that $x=r\cos\varphi$ and $y=r\sin\varphi$ (see Fig.\ \ref{fig:Scheme}). We then get
\begin{equation}
\begin{aligned}
	\vect{m}=
	\begin{bmatrix}
		m_{\rm{r}}(r)\cos\varphi-m_{\rm{\varphi}}(r)\sin\varphi \\
		m_{\rm{\varphi}}(r)\cos\varphi+m_{\rm{r}}(r)\sin\varphi \\
		m_{\rm{z}}
	\end{bmatrix},
\end{aligned}
\end{equation}
\begin{equation}
\begin{aligned}
	\frac{\partial\vect{m}}{\partial x}=
	\begin{bmatrix}
		\cos^2\varphi\frac{\partial m_{\rm{r}}}{\partial r}+\sin^2\varphi\frac{m_{\rm{r}}}{r} \\
		-\sin\varphi\cos\varphi\frac{\partial m_{\rm{\varphi}}}{\partial r}+\sin\varphi\cos\varphi\frac{m_{\rm{\varphi}}}{r} \\
		\cos^2\varphi\frac{\partial m_{\rm{\varphi}}}{\partial r}+\sin^2\varphi\frac{m_{\rm{\varphi}}}{r} \\
		-\sin\varphi\cos\varphi\frac{\partial m_{\rm{r}}}{\partial r}+\sin\varphi\cos\varphi\frac{m_{\rm{r}}}{r} \\
		\cos\varphi m_{\rm{z}}
	\end{bmatrix}
\end{aligned}
\end{equation}
and
\begin{equation}
\begin{aligned}
	\frac{\partial\vect{m}}{\partial y}=
	\begin{bmatrix}
		\sin\varphi\cos\varphi\frac{\partial m_{\rm{r}}}{\partial r}-\sin\varphi\cos\varphi\frac{m_{\rm{r}}}{r} \\
		-\sin^2\varphi\frac{\partial m_{\rm{\varphi}}}{\partial r}-\cos^2\varphi\frac{m_{\rm{\varphi}}}{r} \\
		\sin\varphi\cos\varphi\frac{\partial m_{\rm{\varphi}}}{\partial r}-\sin\varphi\cos\varphi\frac{m_{\rm{\varphi}}}{r} \\
		+\sin^2\varphi\frac{\partial m_{\rm{r}}}{\partial r}+\cos^2\varphi\frac{m_{\rm{r}}}{r} \\
		\sin\varphi m_{\rm{z}}
	\end{bmatrix},
\end{aligned}
\end{equation}
which gives
\begin{equation}
\begin{aligned}
	\interpar{\frac{\partial\vect{m}}{\partial x}\times\frac{\partial\vect{m}}{\partial y}}=
	\begin{bmatrix}
		-\frac{m_{\rm{x}}}{r}\frac{\partial m_{\rm{z}}}{\partial r} \\
		-\frac{m_{\rm{y}}}{r}\frac{\partial m_{\rm{z}}}{\partial r} \\
		-\frac{m_{\rm{z}}}{r}\frac{\partial m_{\rm{z}}}{\partial r} 
	\end{bmatrix}
\end{aligned}
\end{equation}
and then
\begin{equation}
\begin{aligned}
	\vect{G}&=-\frac{M_{\rm{S}}t_{\rm{FM}}}{\gamma}\int_{r=0}^{\infty}\int_{\varphi=0}^{2\pi}\interpar{-\frac{1}{r}\frac{\partial m_z}{\partial r}}rdr \\
	&= \frac{2\pi M_{\rm{S}}t_{\rm{FM}}}{\gamma}\left[ m_z(\infty)-m_z(0)\right].
\end{aligned}
\end{equation}

By keeping only even orders both in $\cos\varphi$ and $\sin\varphi$ (others will integrate to zero), we get
\begin{equation}
	\frac{\partial\vect{m}}{\partial x}\cdot\frac{\partial\vect{m}}{\partial y}=
\interpar{\sin^2\varphi-\cos^2\varphi}\interpar{\frac{\partial m_{\rm{r}}}{\partial r}\frac{m_{\rm{\varphi}}}{r}-\frac{\partial m_{\rm{\varphi}}}{\partial r}\frac{m_{\rm{r}}}{r}}
\end{equation}
and
\begin{equation}
\begin{aligned}
	\frac{\partial\vect{m}}{\partial x}\cdot\frac{\partial\vect{m}}{\partial x}=&\cos^2\varphi \left[ \interpar{\frac{\partial m_{\rm{r}}}{\partial r}}^2+\interpar{\frac{\partial m_{\rm{\varphi}}}{\partial r}}^2+\interpar{\frac{\partial m_{\rm{z}}}{\partial r}}^2 \right] \\
& +\sin^2\varphi\interpar{\frac{1-m_{\rm{z}}}{r^2}}
\end{aligned}
\end{equation}
and we have a similar result with $\partial\vect{m}/\partial y\cdot\partial\vect{m}/\partial y$ so that after integration over $\varphi$ we obtain
\begin{equation}
\begin{aligned}
	\mathcal{D}_{\rm{xx,yy}}&=\frac{\pi M_{\rm{S}}t_{\rm{FM}}}{\gamma}\int_{r=0}^{\infty} \frac{1-m_{\rm{z}}^2}{r}+r\sum_{i=x,\varphi,z}\interpar{\frac{\partial m_{\rm{i}}}{\partial r}}^2 dr \\
	\mathcal{D}_{\rm{xy,yx}}&=0.
\end{aligned}
\end{equation}

The torque exerted by the vertical spin current, polarized along $\vect{y}$, considering only the damping-like term originating in the spin Hall effect of an in-plane current density $J$, is given by
\begin{equation}
	\vect{\Gamma}_{\rm{SOT}}=\frac{\gamma\hbar}{2e\mu_0M_{\rm{S}}t_{\rm{FM}}}\interpar{-\theta_{\rm{SH}}J} \vect{m}\times\interpar{\vect{m}\times\vect{y}}
\end{equation}
with $\theta_{\rm{SH}}$ the effective spin Hall angle of the nearby materials enclosing the ferromagnetic layer. Inserting $\vect{\Gamma}_{\rm{SOT}}$ into \eqref{eq:Thielefactors} we find after integration over $\varphi$
\begin{equation}
\begin{aligned}
	F_{\rm{x}}&=-\theta_{\rm{SH}}J\frac{\pi\hbar}{2e}\int_{r=0}^{\infty}r\interpar{\frac{\partial m_{\rm{r}}}{\partial r}m_{\rm{z}}-\frac{\partial m_{\rm{z}}}{\partial r}m_{\rm{r}}}+m_{\rm{r}}m_{\rm{z}}\>dr \\
	F_{\rm{y}}&=\theta_{\rm{SH}}J\frac{\pi\hbar}{2e}\int_{r=0}^{\infty}r\interpar{\frac{\partial m_{\rm{\varphi}}}{\partial r}m_{\rm{z}}-\frac{\partial m_{\rm{z}}}{\partial r}m_{\rm{\varphi}}}+m_{\rm{\varphi}}m_{\rm{z}}\>dr
\end{aligned}
\end{equation}

\newpage
\bibliography{Skyrmion_profile}

\end{document}